\documentclass[12pt]{article}
\usepackage{epsfig}
\usepackage{times}
\date{}

\begin{document}

\title{{\LARGE\sf Ordering and Broken Symmetry in Short-Ranged Spin Glasses}}
\author{
{\bf C. M. Newman}\thanks{Partially supported by the 
National Science Foundation under grant DMS-01-02587.}\\
{\small \tt newman\,@\,cims.nyu.edu}\\
{\small \sl Courant Institute of Mathematical Sciences}\\
{\small \sl New York University}\\
{\small \sl New York, NY 10012, USA}
\and
{\bf D. L. Stein}\thanks{Partially supported by the 
National Science Foundation under grant DMS-01-02541.}\\
{\small \tt dls\,@\,physics.arizona.edu}\\
{\small \sl Depts.\ of Physics and Mathematics}\\
{\small \sl University of Arizona}\\
{\small \sl Tucson, AZ 85721, USA}
}

\maketitle

\begin{abstract}
In this topical review we discuss the nature of the low-temperature phase
in both infinite-ranged and short-ranged spin glasses.  We analyze the meaning of
pure states in spin glasses, and distinguish between physical, or
``observable'', states and (probably) unphysical, ``invisible'' states.  We
review replica symmetry breaking, and describe what it would mean in
short-ranged spin glasses.  We introduce the notion of thermodynamic chaos,
which leads to the metastate construct.  We apply these tools to short-ranged spin
glasses, and conclude that replica symmetry breaking, in any form, cannot
describe the low-temperature spin glass phase in any finite dimension.  We
then discuss the remaining possible scenarios that {\it could\/} describe
the low-temperature phase, and the differences they
exhibit in some of their physical properties --- in
particular, the interfaces that separate them.  We also
present rigorous results on metastable states and discuss their connection
to thermodynamic states.  Finally, we discuss some of the differences
between the statistical mechanics of homogeneous systems and those with
quenched disorder and frustration.
\end{abstract}

{\bf KEY WORDS:\/} spin glass; Edwards-Anderson model;
Sherrington-Kirkpatrick model; replica symmetry breaking; mean-field
theory; pure states; ground states; metastates; domain walls; interfaces;
metastable states

\vfill\eject

\small
\renewcommand{\baselinestretch}{1.25}
\normalsize

\tableofcontents

\vfill\eject

\section{Introduction}
\label{sec:intro}

Despite decades of intensive investigation, the statistical mechanics
of systems with both quenched disorder and frustration remains an open
problem.  Among such systems, the spin glass is arguably the prototype,
and inarguably the most studied.

\begin{figure}
  \label{fig:ground}
  \centerline{\epsfig{file=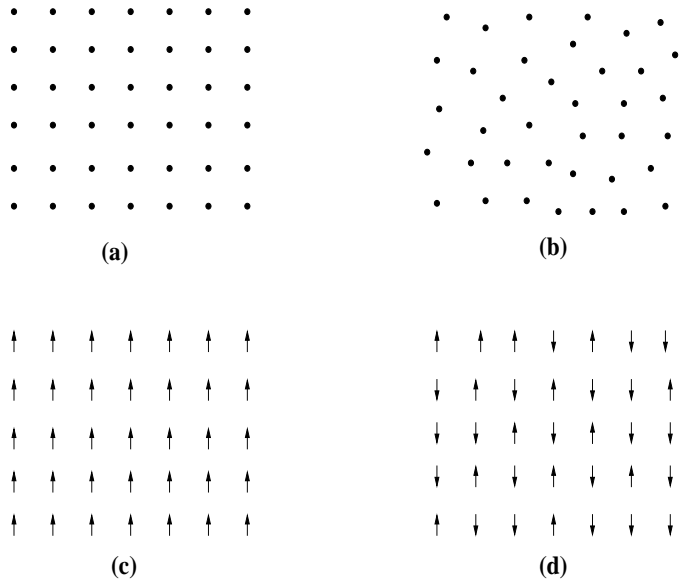,width=3.5in,height=3.0in}}
  \renewcommand{\baselinestretch}{1.0} 
  \small
  \caption{A rough sketch of the classical ground states of: (a) a crystal;
  (b) a glass; (c) a ferromagnet; (d) an Ising spin glass.  In (a) and (b)
  the dots represent atoms; in (c) and (d) the arrows represent localized
  magnetic moments.  (In the case of (b), it is more accurate to describe the
  configuration as a frozen metastable state.)}
  \end{figure}

Spin glasses are systems in which competition between ferromagnetic and
antiferromagnetic interactions among localized magnetic moments (or more
colloquially, ``spins'') leads to a magnetically disordered state
(Fig.~\ref{fig:ground}).  The prime example of a metallic spin glass is a
dilute magnetic alloy, in which a magnetic impurity (typically Fe or Mn) is
randomly diluted within a nonmagnetic metallic host, typically a noble
metal.  The competition between ferromagnetic and antiferromagnetic
interactions arises in these systems from the RKKY interactions
\cite{RK54,K56,Y57} between the localized spins, mediated by the conduction
electrons.
  
  \renewcommand{\baselinestretch}{1.25}
  \normalsize

  But many other types of spin glass, with different microscopic mechanisms
  for their ``spin glass-like'' behavior, exist.  Certain insulators, in
  which low concentrations of magnetic impurities are randomly substituted
  for nonmagnetic atoms, also display spin glass behavior.  A well-known
  example \cite{MF79} is Eu$_x$Sr$_{1-x}$S, with $x$ roughly between .1 and
  .5, where the competition arises largely from nearest-neighbor
  ferromagnetic and next-nearest-neighbor antiferromagnetic interactions.
  There are many other materials that exhibit spin glass behavior, both
  metallic and insulating, crystalline and amorphous.  They can display
  Ising, planar, or Heisenberg behavior, and come in both classical and
  quantum varieties.  In this review we consider only classical spin glasses
  \cite{Bh97}.

  What are the main experimental features of spin glasses?  One is the
  presence of a cusp in the low-field ac susceptibility
  (Fig.~2), as first observed in {\it Au\/}Fe alloys by Cannella
  and Mydosh \cite{CM72}.  This cusp becomes progressively rounded as the
  external magnetic field increases \cite{My87}.  The specific heat, however,
  rather than showing a similar singularity, typically displays a broad
  maximum (Fig.~3) at temperatures somewhat higher than the
  ``freezing temperature'' $T_f$ defined via the susceptibility peak (see,
  e.g., \cite{WK76}).

  \begin{figure}
  \label{fig:susc}
  \centerline{\epsfig{file=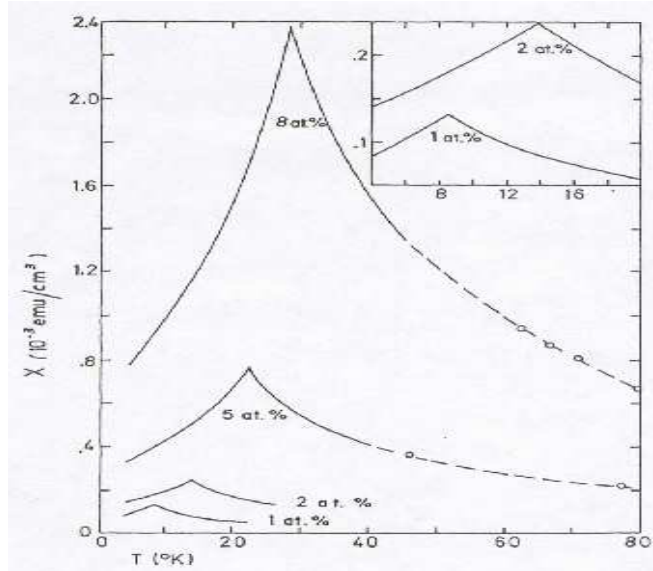,width=3.5in,height=3.0in}}
  \renewcommand{\baselinestretch}{1.0} 
  \small
  \caption{Low-field magnetic susceptibility $\chi(T)$ in {\it Au\/}Fe alloys
  at varying concentrations of iron impurity. From Cannella and Mydosh \cite{CM72}.}
  \end{figure}
  \renewcommand{\baselinestretch}{1.25}
  \normalsize

  \begin{figure}
  \label{fig:Cm}
  \centerline{\epsfig{file=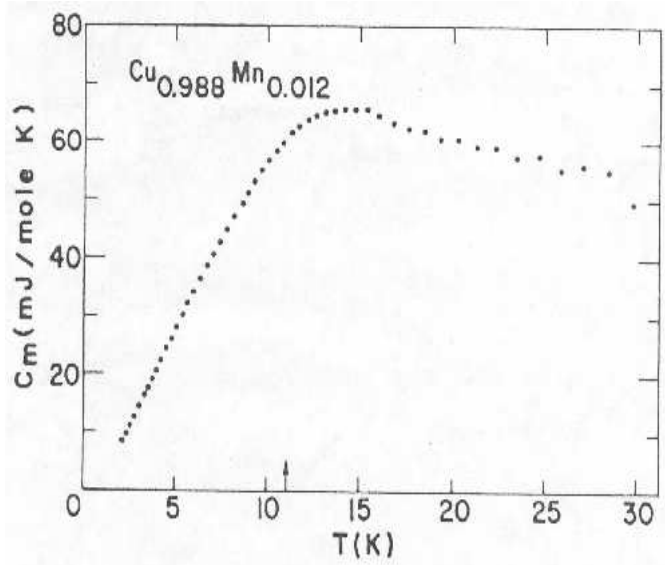,width=3.5in,height=3.0in}}
  \renewcommand{\baselinestretch}{1.0} 
  \small
  \caption{Magnetic specific heat of $C_m$ of {\it Cu\/}Mn at 1.2\% manganese
  impurity.  The arrow indicates the freezing temperature $T_f$ as discussed
  in the text.  From Wenger and Keesom \cite{WK76}.}
  \end{figure}
  \renewcommand{\baselinestretch}{1.25}
  \normalsize

  Probes of the low-temperature magnetic structure using neutron scattering,
  M\"ossbauer studies, NMR and other techniques confirm the absence of
  long-range spatial order coupled with the presence of temporal order
  insofar as the spin orientations appear to be frozen on the timescale of
  the experiment.  An extensive description of these and related experiments
  are presented in the review article by Binder and Young \cite{BY86}.

  Spin glasses are also characterized by a host of irreversible and
  non-equilibrium behaviors, including remanence, hysteresis, anomalously
  slow relaxation, aging and related phenomena.  Because this review will
  focus on static equilibrium behavior, these topics will not be treated
  here, but it is important to note that explaining these phenomena
  are essential to any deeper understanding of spin glasses.  For reviews,
  see \cite{BY86,Chowd86,FH91,BCKM98}.

  The quest to attain a theoretical understanding of spin glasses has
  followed a tortuous path, and to this day many of the most basic and
  fundamental issues remain unresolved.  An extensive discussion of
  theoretical ideas can be found in a number of reviews
  \cite{BY86,Chowd86,FH91,MPV87,Stein89,NSBerlin,Dotsenko01}.  The good news
  is that research into spin glasses has uncovered a variety of novel and
  sometimes stunning concepts; the bad news is that it is not clear how many
  of these apply to real spin glasses themselves.  In this review we will
  explore some of these issues, in particular the nature of ordering and
  broken symmetry in the putative spin glass phase.

  As a first step, one needs to capture mathematically the absence of
  orientational spin ordering in space with the presence of frozenness, or
  order in time.  This was achieved early on by Edwards and Anderson (EA)
  \cite{EA75}, who noted that a low temperature pure phase of spin glasses
  should be characterized by a vanishing magnetization per spin
  \begin{equation}
  \label{eq:mag}
  M=\lim_{L\to\infty}{1\over|\Lambda_L|}\sum_{x\in\Lambda_L}\langle\sigma_x\rangle
  \end{equation}
  accompanied by a nonvanishing ``EA order parameter'' (as it is now called)
  \begin{equation}
  \label{eq:qea}
  q_{EA}=\lim_{L\to\infty}{1\over|\Lambda_L|}\sum_{x\in\Lambda_L}\langle\sigma_x\rangle^2
  \end{equation}
  where $\sigma_x$ is the spin at site $x$, $\Lambda_L$ is a cube of side $L$
  centered at the origin, and $\langle\cdot\rangle$ denotes a thermal average.

  However, it was later discovered that, while the EA order parameter
  $q_{EA}$ plays a central role in describing the spin glass phase, it is
  insufficient to describe the low temperature ordering --- at least in a
  mean-field version of the problem.  In the following sections we will
  explain this statement, explore the relationship between the mean-field
  spin glass problem and its short-ranged version, and discuss some new and
  general insights and tools that may turn out to be useful in unraveling the
  complexities of the statistical mechanics of inhomogeneous systems.

  \section{A Brief History of Early Theoretical Developments}
  \label{sec:history}

  Questions regarding spin glass behavior fall naturally into two classes:
  the first pertains to properties of a system in thermal equilibrium, and
  the second to those related to nonequilibrium dynamics.  It is still not
  certain whether spin glasses possess nontrivial equilibrium properties, but
  their nonequilibrium ones surely are.  As already noted, throughout this
  review we will focus primarily on spin glasses in thermal equilibrium.
   
   Perhaps the most fundamental question that can be asked is whether there
   exists such a thing as a true, thermodynamic spin glass phase; that is, is
   there a sharp phase transition from the high-temperature paramagnetic state
   to a low-temperature spin glass state in zero external magnetic field?  Of
   course, this is presumably the simplest transition that could in principle
   occur; one could also ask about field-induced transitions, ferromagnetic to
   spin glass transitions, and others.  But given that all of these questions
   remain open, we'll confine our attention here to the simplest of these.
   And even here, although experimental and numerical studies have tended to
   favor an affirmative answer, the issue is by no means resolved.

   \subsection{The Edwards-Anderson Hamiltonian}
   \label{subsec:EA}

   In order to proceed, we need a specific model to study.  The majority of
   theoretical investigations begin with a Hamiltonian proposed by Edwards and
   Anderson \cite{EA75}:
   \begin{equation}
   \label{eq:EA}
   {\cal H}=-\sum_{<x,y>} J_{xy} \sigma_x\sigma_y -h\sum_x\sigma_x\ ,
   \end{equation}
   where (to keep things as general as possible) $x$ is a site in a
   $d$-dimensional cubic lattice, $\sigma_x$ is the spin at site $x$, the spin
   couplings $J_{xy}$ are independent, identically distributed random
   variables, $h$ is an external magnetic field, and the first sum is over
   nearest neighbor sites only.  We will usually take the spins to be Ising
   variables, i.e., $\sigma_x=\pm 1$.  Throughout most of the paper we will
   choose $h=0$ and the spin couplings $J_{xy}$ to be symmetrically
   distributed about zero; as a result, the EA Hamiltonian in
   Eq.~(\ref{eq:EA}) has global spin inversion symmetry.  

   Popular choices for the distribution of the couplings $J_{xy}$ are bimodal
   and Gaussian.  Most of what we discuss below will be independent of which
   of these is chosen, but for specificity (and to avoid accidental
   degeneracies when discussing ground states) we will choose the couplings
   from a Gaussian distribution with mean zero and variance one.  It is
   important to note the the disorder is {\it quenched\/}: once chosen, the
   couplings are fixed for all time.  We denote by ${\cal J}$ a particular
   realization of the couplings, corresponding physically to a specific spin
   glass sample.  Proper averaging over quenched disorder is done on extensive
   quantities only \cite{Br59}, i.e., at the level of $\log Z$ rather than
   $Z$, where $Z$ is the partition function.

   Of course, the EA Hamiltonian looks nothing like a faithful microscopic
   description of spin interactions in a dilute magnetic alloy, or an
   insulator like Eu$_x$Sr$_{1-x}$S --- and because the statistical mechanics
   of the EA Hamiltonian remain to be worked out, a direct comparison with
   experiment remains elusive.  However, a central assertion of \cite{EA75} is
   that the essential physics of spin glasses is the competition between
   quenched ferromagnetic and antiferromagnetic interactions, regardless of
   microscopic details; and so the EA Hamiltonian remains the usual launching
   point for theoretical analyses of real spin glasses.

   \subsubsection{Frustration}
   \label{subsec:frustration}

   A striking feature of random-bond models like the EA spin glass is the
   presence of {\it frustration\/}, in this case meaning the inability of any
   spin configuration to simultaneously satisfy all couplings.  It is easily
   verified that, in any dimension larger than one, all of the spins along any
   closed circuit ${\cal C}$ in the edge lattice cannot be simultaneously
   satisfied if
   \begin{equation}
   \label{eq:frustration}
   \prod_{<x,y>\in{\cal C}}J_{xy}<0\, .
   \end{equation}

   The definition of frustration given above was first suggested by Toulouse
   \cite{T77}.  A different formulation due to Anderson \cite{And78} received
   less notice when it was first proposed, but its underlying ideas may prove
   useful in more recent spin glass research.  The basic notion is that
   frustration manifests itself as free energy fluctuations (e.g., with a
   change in boundary conditions from periodic to antiperiodic) that scale as
   the {\it square root\/} of the surface area of a typical sample.

   Hence the spin glass is characterized by both {\it quenched disorder\/} and
   {\it frustration\/}.  The presence of frustration, leading to a complicated
   geometry of entangled frustration contours, suggests the possibility that
   spin glasses, in at least some dimensions, may possess multiple
   infinite-volume ground or pure states unrelated by any simple symmetry
   transformation.  We will return to this question later.  We note here,
   though, that there exists at least one (unrealistic) spin glass model where
   the number of ground states can be computed in all finite dimensions.  This
   is the {\it highly disordered model\/} of the authors \cite{NS94,NS96a}
   (see also \cite{BCM94}) in which the coupling magnitudes scale nonlinearly
   with the volume (and so are no longer distributed independently of the
   volume, although they remain independent and identically distributed for
   each volume).  It can be shown \cite{NS94,NS96a} that there exists a {\it
   transition\/} in ground state multiplicity in this model: below eight
   dimensions, it has only a single (globally spin reversed) pair of ground
   states, while above eight it has uncountably many ground state pairs.
   Interestingly, the high-dimensional ground state multiplicity can be shown
   to be {\it unaffected\/} by the presence of frustration, although
   frustration still plays an interesting role: it leads to the appearance of
   {\it chaotic size dependence\/}, to be discussed in Sec.~\ref{sec:csd}.

   \subsection{Mean Field Theory, the Sherrington-Kirkpatrick Model, and the
   Parisi Solution}
   \label{subsec:SK}

   Mean field models often provide a useful first step towards understanding
   the low-temperature phase of a condensed matter system; in the case of spin
   glasses, the usual procedure seems to have taken a particularly interesting
   twist.  The mean field theory of spin glasses turns out to be far more
   intricate than those of most homogeneous systems, and as a result several
   different approaches have been tried.  Also noteworthy are even simpler,
   soluble spin glass-like models, in particular the random energy model of
   Derrida \cite{Derr80}.  However, here we will confine ourselves to a
   discussion of the Sherrington-Kirkpatrick (SK) model \cite{SK75}, an
   infinite-ranged version of the EA model in which mean field theory is
   presumably exact.

   The SK Hamiltonian for a system of $N$ spins is (as usual, we take external
   field to be zero):
   \begin{equation}
   \label{eq:SK}
   {\cal H}_N=-(1/\sqrt{N})\sum_{1\le i<j\le N} J_{ij} \sigma_i\sigma_j 
   \end{equation}
   where we again take the spins to be Ising and choose the (independent,
   identically distributed) couplings $J_{ij}$ from a Gaussian distribution
   with zero mean and variance one; the latter rescaling ensures a sensible
   thermodynamic limit for free energy per spin and other thermodynamic
   quantities.

   It was shown in \cite{SK75} that this model has a sharp phase transition at
   $T_c=1$.  At this temperature the static susceptibility has a cusp --- but
   so does the specific heat.

   SK solved for the low-temperature spin glass phase, using the EA order
   parameter to describe the broken symmetry.  However, their solution was
   unstable \cite{SK75}; in particular, the entropy was found to become negative
   at sufficiently low temperature.

   The following four years saw intensive efforts to solve for the
   low-temperature phase of the SK model.  Of particular note is the direct
   mean field approach of Thouless, Anderson, and Palmer \cite{TAP77}, who
   pointed out the necessity of including the Onsager reaction field term,
   and the paper by deAlmeida and Thouless \cite{AT78}, who studied the
   stability of the SK solution in the $T$-$h$ plane and calculated the
   boundary between the regions where a single (i.e., paramagnetic) phase
   is stable and the region where the low-temperature phase resides.  One
   important question that remains open to this day is whether such an ``AT
   line'' exists for more realistic models (see
   Sec.~\ref{subsec:magnetic} for further discussion).

   We will mainly focus, however, on what is today believed to be the correct
   solution for the low-temperature phase of the SK model.  In a series of
   papers, Parisi and collaborators \cite{P79,P83,MPSTV84a,MPSTV84b} proposed,
   and worked out the consequences of, an extraordinary {\it ansatz\/} for the
   nature of this phase.  Following the mathematical procedures underlying the
   solution, it came to be known as {\it replica symmetry breaking\/} (RSB).

   We will not review those mathematical procedures here; they are worked out
   in detail in several review articles and books (see, e.g.,
   \cite{BY86,Chowd86,FH91,MPV87,Stein89,Dotsenko01}).  We will also omit
   discussion of important related developments, such as the dynamical
   interpretation of Sompolinsky and Zippelius \cite{SZ81,Somp81,SZ82}.  We
   will concern ourselves instead with both the physical and mathematical
   interpretations of the Parisi solution, and the type of ordering that it
   implies.

   These interpretations took several years to work out, culminating in the
   work of Mezard {\it et al.\/} \cite{MPSTV84a,MPSTV84b} that introduced the
   ideas of overlaps, non-self-averaging, and ultrametricity as a means of
   understanding the type of order implied by the Parisi solution.  These
   terms, and their relevance for the Parisi solution, will be described in
   Sec.~\ref{sec:mft}.  For now, we simply note that the solution of the
   infinite-ranged SK model generated tremendous excitement; as described by
   Rammal {\it et al.\/} \cite{RTV86}, it displayed a new type of broken
   symmetry ``radically different from all previously known''.  This is not an
   overstatement.

   The starting point is the observation that the low-temperature phase
   consists not of a single spin-reversed pair of states, but rather
   ``infinitely many pure thermodynamic states'' \cite{P83}, not related by
   any simple symmetry transformations.  This possibility had already been
   foreshadowed by the Thouless-Anderson-Palmer approach \cite{TAP77}, whose
   mean-field equations were known to have many solutions (not necessarily all
   free energy minima, except at zero temperature \cite{BM80}).  The existence
   of many states meant that the correct order parameter needed to reflect
   their presence, and to describe the relations among them.  The single EA
   order parameter was therefore insufficient to describe the low-temperature
   phase (although it retained an important role, as we'll see); instead one
   needed an order parameter {\it function\/}.

   Before we describe these ideas in more detail, we will first step back and
   consider the basic outline of the problem that interests us.  In
   particular, we need to ask:  what is it that we want to know?  What
   are the fundamental open questions?  And how do they tie in with the
   broader areas of condensed matter physics and statistical mechanics?
   These questions will be considered in the following section.

   \section{Open Problems}
   \label{sec:open}

   In this review we concern ourselves with perhaps the most basic questions
   that can be asked: is there a true spin glass phase, and if so, what is its
   nature?

   For the infinite-ranged SK model, these questions appear largely resolved,
   though some open questions remain.  For the EA model (as a representative
   of more general ``realistic'' models, i.e., finite-dimensional and
   non-infinite-ranged), the primary question of whether a thermodynamic phase
   transition exists remains open.  There is suggestive analytical
   \cite{FS90,TH96} and numerical \cite{BY86,O85,OM85,KY96} evidence that a
   phase transition to a broken spin-flip symmetric phase is present in
   three-dimensional and, even more likely, in four-dimensional Ising spin
   glasses.  However, no one has yet been able to prove or disprove the
   existence of a phase transition, and the issue remains unsettled
   \cite{MPR94}.

   Of course, existence of a phase transition does not necessarily imply more
   than a single low-temperature phase; one could, for example, have a
   transition above which correlations decay exponentially and below which
   they decay as a power law, with $q_{EA}=0$ at all nonzero temperatures.
   However, most numerical simulations and theoretical pictures that point to
   a low-temperature spin glass phase suggest broken spin-flip symmetry.  We are
   therefore led to:

   \smallskip

   {\em Open Question 1.\/} Does the EA Ising model have an equilibrium phase
   transition above some lower critical dimension $d_c$; and if so, does the
   low-temperature phase have broken spin-flip symmetry?

   \smallskip

   If the answer to this question turns out to be no, then subsequent research
   will need to focus on dynamical behavior, which --- as in ordinary glasses
   --- presents a range of difficult and important problems.  However, given
   the reasonable possibility that there is indeed a sharp phase transition,
   it is worthwhile to ask:

   \smallskip

   {\sl If there is a phase transition for the EA Ising model at some $d>d_c$,
   what is the nature of the ordering of the low-temperature phase?\/}

   \smallskip

   Because of the open-ended nature of this question, it won't be assigned a
   number; instead, we'll break it down into several parts.  The remaining
   questions assume that there is an equilibrium phase transition
   critical temperature $T_c>0$, below which there is broken spin-flip symmetry
   (equivalently, a phase with $q_{EA}>0$), but we make no assumptions as to
   whether $d_c$ is less than or equal to three.

   \smallskip

   {\em Open Question 2.\/} What is the number of equilibrium pure state pairs
   (at nonzero temperature) and ground state pairs (at zero temperature) in
   the spin glass phase?

   \smallskip

   We have seen that the mean-field RSB picture assumes infinitely many such
   pairs.  A competing picture, known as the droplet/scaling picture, due to
   Macmillan, Bray and Moore, Fisher and Huse, and others
   \cite{Mac84,BM85,BM87,FH86,HF87a,FH87b,FH88}, asserts that there is only a
   single pair of pure/ground states in the spin glass phase in {\it any\/}
   finite dimension. 

   Because of the importance of this picture, we discuss it briefly here.
   ``Domain wall renormalization group'' studies \cite{Mac84,BM85} led to a
   scaling ansatz \cite{Mac84,BM85,FH86} that in turn led to the development
   of a corresponding physical droplet picture \cite{FH86,HF87a,FH87b,FH88}
   for spin glasses.  In this picture, thermodynamic and dynamic properties at
   low temperature are dominated by low-lying excitations corresponding to
   clusters of coherently flipped spins.  The density of states of these
   clusters at zero energy falls off as a power law in lengthscale $L$, with
   exponent bounded from above by $(d-1)/2$.  At low temperatures and on large
   lengthscales the thermally activated clusters form a dilute gas and can be
   considered as non-interacting two-level systems.  The resulting two-state
   picture (in which there is no nontrivial replica symmetry breaking) is
   therefore significantly different from the mean-field picture arising from
   the SK model.

   So do spin glasses in finite dimensions have many equilibrium states or a
   single pair?  Except in the highly disordered model \cite{NS94,NS96a}, the
   answer is not known.  In one dimension (where there is no internal
   frustration), there is only a single pair of ground states, and a single
   paramagnetic phase at all nonzero temperatures.  In an infinite number of
   dimensions, there presumably would be an infinite number of pure state
   pairs for $T<T_c$.  Recent numerical experiments \cite{Mid99,PY99a,Hart99}
   {\it seem\/} to indicate a single pair of ground states in two dimensions
   (where it is believed that $T_c=0$), but given that lattice sizes studied
   are still not very large, the question is not completely settled.  Recent
   rigorous work by the authors \cite{NS00,NS01a} has led to a partial result
   that supports the notion that only a single pair of ground states occurs in
   two dimensions.  In three dimensions numerical simulations give conflicting
   results \cite{PY99b,MP00}.

   While the mean-field-like RSB many-state picture and the two-state
   droplet/scaling picture have historically been the main competing pictures,
   there are others as well.  At least one of these will be discussed later.
   One often sees in the literature an unspoken assumption that the presence
   of many states is synonymous with RSB, and similarly that the presence of
   only a single pair is equivalent to droplet/scaling.  We emphasize,
   however, that while these are {\it necessary\/} requirements for each
   picture, respectively, they are not sufficient: each has considerable
   additional structure (which, in the mean-field RSB case, will be discussed
   in the next section).  This then leads to our next question:

   \smallskip

   {\em Question 3.\/}  If there do exist infinitely many equilibrium
   states in some dimensions and at some temperatures, are they organized
   according to the mean-field RSB picture? 

   \smallskip

   Treatment of this question is the main theme of the remainder of this
   review.  A series of both rigorous and heuristic results, due to the
   authors, has largely answered this question in the negative, and it is
   therefore not listed as open.  (A complete discussion is given in
   Sec.~\ref{sec:mfscenario}.)  However, there are remaining questions, such
   as:

   \smallskip

   {\em (Semi-)Open Question 4.\/} What are the remaining possibilities for
   the number and organization of equilibrium states in the low-temperature
   spin glass phase? 
   This question is examined in Sec.~\ref{subsec:possibilities}.

   \smallskip

   In discussing this, we will not consider every logical alternative to the
   mean-field picture, but rather what we consider to be the most likely
   remaining scenarios for the low-temperature phase of finite-dimensional
   spin glasses.

   The discussion so far has considered only equilibrium pure or ground
   states, with a view towards determining the nature of broken symmetry in
   realistic spin glasses. However, a more general discussion of
   thermodynamics and dynamics, particularly with a view towards explaining
   experimental observations, needs to include questions about other types of
   states, such as:

   \smallskip

   {\em Question 5.\/} How are energetically low-lying excitations above the
   ground state(s) characterized? (Sec.~\ref{subsec:excitations}.)

   \smallskip

   {\em Question 6.\/} What can be proven about numbers and overlaps of
   metastable states?  Do they have any connection(s) to thermodynamic pure
   states?  (Sec.~\ref{subsec:metastable}.)

   \smallskip

   Although many other important questions remain open, we close here with a
   question of more general interest than for spin glass physics alone:

   \smallskip

   {\em Question 7.\/} In what ways do we now understand how the statistical
   mechanical treatment of frustrated, disordered systems differs in
   fundamental ways from that of homogeneous systems?
   (Sec.~\ref{subsec:statmech}.)

   \section{Nature of Ordering in the Infinite-Ranged Spin Glass}
   \label{sec:mft}

   We now return to a more detailed discussion of the nature of ordering
   implied by Parisi's solution of the SK model.  As noted in
   Sec.~\ref{subsec:SK}, the RSB scheme introduced by Parisi assumes the
   existence of many equilibrium pure states.  Because the notion of pure
   states has generated some confusion in the literature, we detour to clarify
   exactly what is meant by this and related terms.  The discussion here
   closely follows Appendix~A of \cite{NS02}.

   \subsection{Thermodynamic Pure States}
   \label{subsec:pure}

   The notion of pure states is well-defined for short-ranged,
   finite-dimensional systems, but is less clear for infinite-ranged ones like
   the SK model.  We therefore begin with a discussion of the EA model (in
   arbitrary $d<\infty$), and then briefly discuss application of these ideas
   to the SK model.

   Consider first ${\cal H}_{J,L}$, the EA Hamiltonian~(\ref{eq:EA}) restricted to a finite
   volume of linear extent $L$.  We will always take such a volume, hereafter
   denoted as $\Lambda_L$, to be an $L^d$ cube centered at the origin. 
   In addition, we need to impose boundary conditions, which we will often
   take to be periodic; other possibilities include antiperiodic, free, fixed
   (e.g., all spins on the boundary set equal to $+1$), and so on.  Given a
   specified boundary condition, the
   finite-volume Gibbs state $\rho_{{\cal J},T}^{(L)}$ on $\Lambda_L$ at
   temperature $T$ is defined by:
   \begin{equation}
   \label{eq:finite}
   \rho_{{\cal J},T}^{(L)}(\sigma)=Z_{L,T}^{-1} \exp \{-{\cal H}_{{\cal
   J},L}(\sigma)/k_BT\}\quad ,
   \end{equation}
   where the partition function $Z_{L,T}$ is such that the sum of
   $\rho_{{\cal J},T}^{(L)}$ over all spin configurations in $\Lambda_L$
   yields one.

   All equilibrium quantities of interest can be computed from $\rho_{{\cal
   J},T}^{(L)}(\sigma)$, which is simply a probability measure: it describes
   at fixed $T$ the probability of a given spin configuration obeying the
   specified boundary condition appearing within $\Lambda_L$.  Such a
   (well-behaved) probability distribution is completely specified by its
   moments, which in this case is the set of all correlation functions within
   $\Lambda_L$: $\langle\sigma_{x_1}\cdots\sigma_{x_m}\rangle$ for arbitrary
   $m$ and arbitrary $x_1,\ldots,x_m\in\Lambda_L$.

   Consider next the $L\to\infty$ limit of a sequence of such finite-volume
   Gibbs states $\rho_{{\cal J},T}^{(L)}(\sigma)$, each with a specified
   boundary condition (which may remain the same or may change with $L$).  Of
   course, such a sequence may or may not have a limit; existence of a limit
   would require that every $m$-spin correlation function, for $m=1,2,\ldots$,
   itself possesses a limit \cite{Geo88}.  A {\it thermodynamic\/} state
   $\rho_{{\cal J},T}$ is therefore an {\it infinite\/}-volume Gibbs measure,
   providing information such as the probability of any finite subset of spins
   taking on specified values; and of course it determines global properties
   such as magnetization per spin, energy per spin, and so on.

   Thermodynamic states may or may not be mixtures of other thermodynamic
   states.  If a thermodynamic state $\rho_{{\cal J},T}$ can be decomposed
   according to
   \begin{equation}
   \label{eq:mixed}
   \rho_{{\cal J},T}=\lambda\rho^{1}_{{\cal J},T}+(1-\lambda)\rho^{2}_{{\cal J},T}\quad ,
   \end{equation}
   where $0 < \lambda < 1$ and $\rho^{1}$ and $\rho^{2}$ are also
   thermodynamic states (distinct from $\rho$), then $\rho_{{\cal J},T}$ is a
   {\it mixed\/} thermodynamic state or simply, mixed state. The meaning of
   the decomposition in Eq.~(\ref{eq:mixed}) is easily understood: it simply
   means that any correlation function computed using $\rho_{{\cal J},T}$ can
   be decomposed as:
   \begin{equation}
   \label{eq:decomposed}
   \langle\sigma_{x_1}\cdots\sigma_{x_m}\rangle_{\rho_{{\cal J},T}} = 
   \lambda\langle\sigma_{x_1}\cdots\sigma_{x_m}\rangle_{\rho^{1}_{{\cal
   J},T}}
   +(1-\lambda)\langle\sigma_{x_1}\cdots\sigma_{x_m}
   \rangle_{\rho^{2}_{{\cal J},T}}\quad .
   \end{equation}
   A mixed state may or may not be further decomposed into as many as an
   uncountable infinity of disjoint other states.

   We are now ready to define the idea of a thermodynamic pure state.  If a
   distinct thermodynamic state cannot be written as a convex combination of
   any other thermodynamic states, it is then a thermodynamic {\it pure\/}
   state, or simply pure state.  So the paramagnetic state is a pure state, as
   are each of the positive and negative magnetization states in the Ising
   ferromagnet.  In that system at $T<T_c$, the Gibbs state produced by a
   sequence of increasing volumes, with either periodic or free boundary
   conditions, is a mixed state, decomposable into the positive and negative
   magnetization states each with $\lambda=1/2$.  On the other hand, a
   sequence of increasing volumes with all boundary spins fixed at $+1$ leads
   to the positive magnetization pure state.

   A thermodynamic, or infinite-volume, {\it ground\/} state is a pure state
   at zero temperature consisting of a single spin configuration, with the
   property that its energy cannot be lowered by flipping any finite subset of
   spins \cite{noteground}.

   A pure state $\rho_P$ can be intrinsically characterized by a
   {\it clustering property\/} (see, e.g., \cite{Geo88,vEvH84}), which implies
   that for any fixed $x$,
   \begin{equation}
   \label{eq:clustering}
   \langle\sigma_x\sigma_y\rangle_{\rho_P} - 
   \langle\sigma_x\rangle_{\rho_P}\langle\sigma_y\rangle_{\rho_P}\ \to 0 ,
   \qquad \vert y \vert\to\infty\quad ,
   \end{equation}
   and similar clustering for higher order correlations.

   Without getting into technical details, it is already clear that a problem
   exists with extending the concept of pure (or for that matter
   thermodynamic) state to the SK model.  Such states are defined for a fixed
   realization ${\cal J}$ of {\it all\/} the couplings; but in the SK model
   the couplings all scale to zero as $N\to\infty$.  However, one can still
   talk in some rough sense about an SK state ${\cal P}$ with a ``modified''
   clustering property
   \begin{equation}
   \label{eq:modclustering}
   \langle\sigma_i\sigma_j\rangle_{\rho_{\cal P}} - 
   \langle\sigma_i\rangle_{\rho_{\cal }P}\langle\sigma_j\rangle_{\rho_{\cal P}}\ \to 0 ,
   \qquad N\to\infty\quad ,
   \end{equation}
   for any pair $i$ and $j$ \cite{BY86}.  The problem with this is that there
   may be no measurable (i.e., effective) way to construct such a state (see
   Sec.~\ref{sec:csd} below).

   For now, we will ignore such difficulties and use the terms ``pure state'',
   ``thermodynamic state'', and so on, for the SK model also, as has been done
   extensively in the physics literature.  It should always be kept in mind,
   though, that such usage is rough, and any attempt to make the notion more
   precise runs into serious difficulties (although we will propose a
   heuristic method for detecting the existence of many states in the SK model
   in Sec.~\ref{sec:csd}).  Note that some of the difficulties discussed here
   are not present in the Curie-Weiss model of the uniform ferromagnet.
   Although couplings scale to zero there also, they ``reinforce'' each other,
   being nonrandom, so that $N\to\infty$ positive and negative magnetization
   states still make sense.  It turns out that this fundamental difference
   between finite and infinite systems in the {\it disordered\/} case will
   have profound consequences.

   \subsection{Overlap Functions and Distributions}
   \label{subsec:overlaps}

   As noted in Sec.~\ref{subsec:SK}, Parisi's mathematical replica symmetry
   breaking scheme led to the physical interpretation of many pure states
   below $T_c$.  All such states at fixed $T$ have vanishing magnetization per
   spin as $N$ gets large, and all have the same (nonzero) $q_{EA}(T)$
   \cite{BY86,MPV87}, so these alone are insufficient to describe the
   ordering.  What is needed is a way to describe the relations among the
   different states, and this can be accomplished by means of {\it
   overlaps\/}.

   The usual interpretation of the Parisi solution is as follows
   \cite{BY86,MPV87,MPSTV84a,MPSTV84b}.  For large $N$, the Gibbs distribution
   is a mixture of many pure states:
   \begin{equation}
   \label{eq:SKdecomp}
   \rho_{{\cal J},N}(T)\approx\sum_\alpha W_{\cal J}^{\alpha}(T)\rho_{\cal
   J}^\alpha(T)\, ,
   \end{equation}
   where $\rho_{\cal J}^\alpha$ is pure state $\alpha$, $W_{\cal J}^{\alpha}$
   its weight in the decomposition of $\rho_{{\cal J},N}(T)$, and the
   approximate equality sign indicates that the notion of pure state isn't
   precise in this model.  Although Eq.~(\ref{eq:SKdecomp}) involves in
   principle an infinite sum, most weights are vanishingly small; only $O(1)$
   states have appreciable weights as $N\to\infty$.

   The overlap between state $\alpha$ and $\beta$ is defined as
   \begin{equation}
   \label{eq:qabSK}
   q_{\alpha\beta}\approx {1\over
   N}\sum_{i=1}^N\langle\sigma_i\rangle_\alpha\langle\sigma_i\rangle_\beta\, ,
   \end{equation}
   where $\langle\cdot\rangle_\alpha$ is a thermal average in pure state
   $\alpha$, and dependence on ${\cal J}$ and $T$ has been suppressed.  So
   $q_{\alpha\beta}$ is a measure of the similarity between states $\alpha$
   and $\beta$.  It is clear that
   \begin{equation}
   \label{eq:qabrange}
   -q_{EA}\le q_{\alpha\beta}\le q_{EA}
   \end{equation}
   because $q_{EA}=q_{\alpha\alpha}$ and $-q_{EA}=q_{\alpha,-\alpha}$, where
   $-\alpha$ is the global flip of $\alpha$ (i.e., all odd-spin correlation
   functions in $\alpha$ and -$\alpha$ have the same magnitude and opposite
   sign, and all even-spin correlations in the two are equal).

   Because there is no spatial structure in the infinite-ranged model, the
   overlap function does seem to capture the essential relations among the
   different states.  However, it might already be noticed that such a global
   measure may well miss important information in short-ranged models ---
   assuming that such models also have many pure states.

   Quantities referring to individual pure states are problematic, since there
   is no known procedure for constructing such things in the SK model.
   However, what is really of interest is the {\it distribution\/} of
   overlaps; i.e., if one randomly and independently chooses two pure states
   from $\rho_{{\cal J},N}(T)$ in Eq.~(\ref{eq:SKdecomp}), the probability
   that their overlap lies between $q$ and $q+dq$ is $P_{\cal J}(q)dq$, where
   \begin{equation}
   \label{eq:ovdistSK}
   P_{\cal J}(q)=\sum_\alpha\sum_\beta W_{\cal J}^{\alpha}W_{\cal
   J}^{\beta}\delta(q-q_{\alpha\beta})\, .
   \end{equation}
   As before, we suppress the dependence on $T$ and $N$ for ease of notation.
   $P_{\cal J}(q)$ is commonly referred to as the {\it Parisi overlap
   distribution\/}.

   What does $P_{\cal J}(q)$ look like?  When there is a single pure state
   with zero magnetization per spin, such as the paramagnet, it is simply a
   $\delta$-function at $q=0$.  For the Curie-Weiss Ising ferromagnet below
   $T_c$ (as well as in short-ranged ferromagnets with, say, periodic boundary
   conditions), there are two pure states: the uniform positive and negative
   magnetization states, each appearing with weight $1/2$ in a pure state
   decomposition of the type described in Eq.~(\ref{eq:SKdecomp}).  The
   resulting overlap distribution is shown in Fig.~4.

   \begin{figure}
   \label{fig:feroverlap}
   \centerline{\epsfig{file=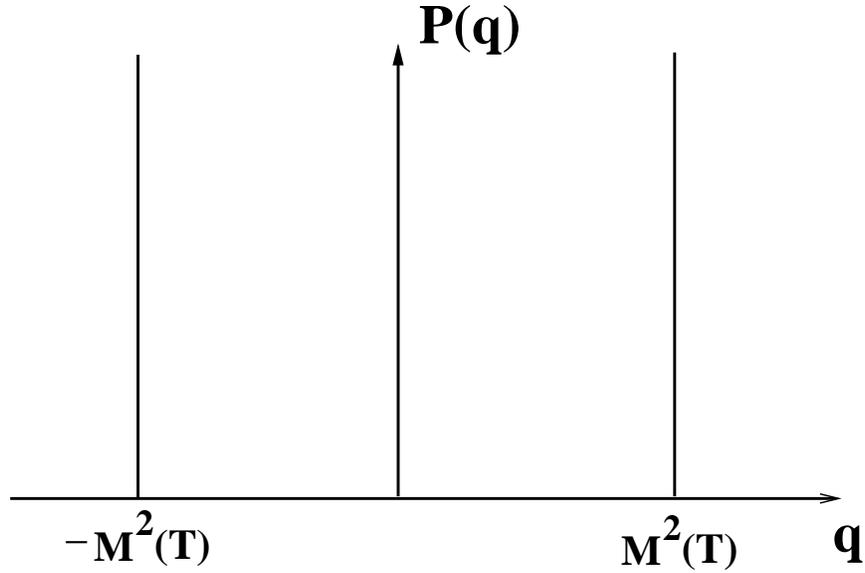,width=4.5in,height=3.0in}}
   \renewcommand{\baselinestretch}{1.0} 
   \small
   \caption{Overlap distribution function $P(q)$ for the Curie-Weiss Ising
   ferromagnet below $T_c$, or for short-ranged ferromagnets with periodic
   boundary conditions.  In this figure, $M(T)$ is the magnetization per spin
   and the spikes at $\pm M^2(T)$ are $\delta$-functions.}
   \end{figure}
   \renewcommand{\baselinestretch}{1.25}
   \normalsize

   For the SK model, the overlap distribution for fixed ${\cal J}$ is
   nontrivial, due to the presence of many non-symmetry-related pairs of
   states in the decomposition of the Gibbs distribution $\rho_{\cal J}$
   (cf.~Eq.~(\ref{eq:SKdecomp})).  Now there are several $\delta$-functions of
   nontrivial weight, distributed symmetrically about zero in the interval
   $[-q_{EA},q_{EA}]$, as sketched in Fig.~5.

   \begin{figure}
   \label{fig:SKoverlap}
   \centerline{\epsfig{file=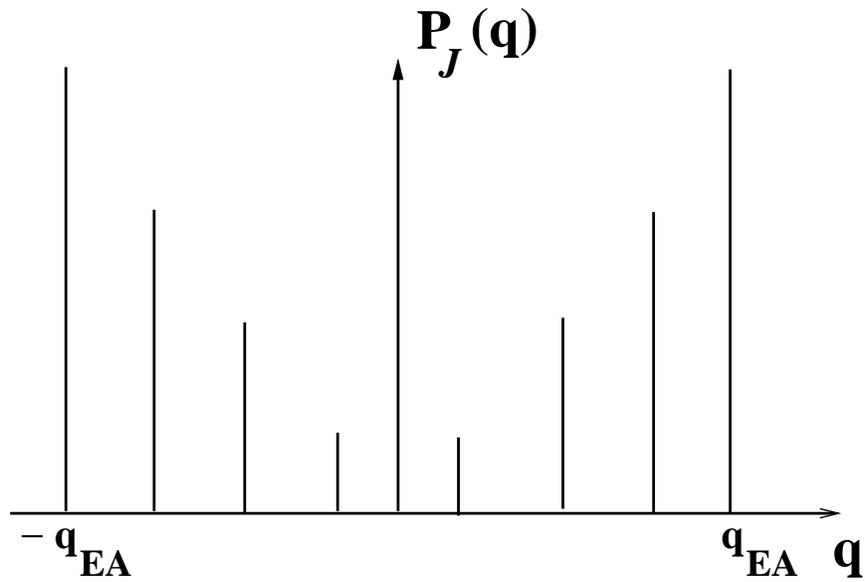,width=4.5in,height=3.0in}}
   \renewcommand{\baselinestretch}{1.0} 
   \small
   \caption{Sketch of the overlap distribution function $P_{\cal J}(q)$ for
   the SK model below $T_c$.}
   \end{figure}
   \renewcommand{\baselinestretch}{1.25}
   \normalsize

   \subsection{Non-Self-Averaging}
   \label{subsec:nsa}

   One of the most interesting and peculiar features of the Parisi solution is
   the {\it non-self-averaging\/} of the overlap distribution function.  For
   large $N$ and fixed ${\cal J}$, $P_{\cal J}(q)$ has the form shown in
   Fig.~5.  What happens if one looks at a ${\cal J'}$ different from ${\cal
   J}$?  Surprisingly, the overlaps (except for the two at $\pm q_{EA}$, which
   are present for almost every ${\cal J}$) will generally appear at different
   values of $q$, and the set of corresponding weights will also differ.  This
   is true no matter how large $N$ becomes.  If one then averages $P_{\cal
   J}(q)$ over all ${\cal J}$ as $N\to\infty$, the resulting distribution will
   be supported on all values of $q$ in the interval $[-q_{EA},q_{EA}]$.

   Let $P_N(q)=\overline{P_{{\cal J},N}(q)}$, where the overbar indicates a
   quenched average over coupling realizations ${\cal J}$, and let
   $P(q)=\lim_{N\to\infty}P_N(q)$.  A sketch of the averaged overlap
   distribution $P(q)$ is shown in Fig.~6.

   \begin{figure}
   \label{fig:Pq}
   \centerline{\epsfig{file=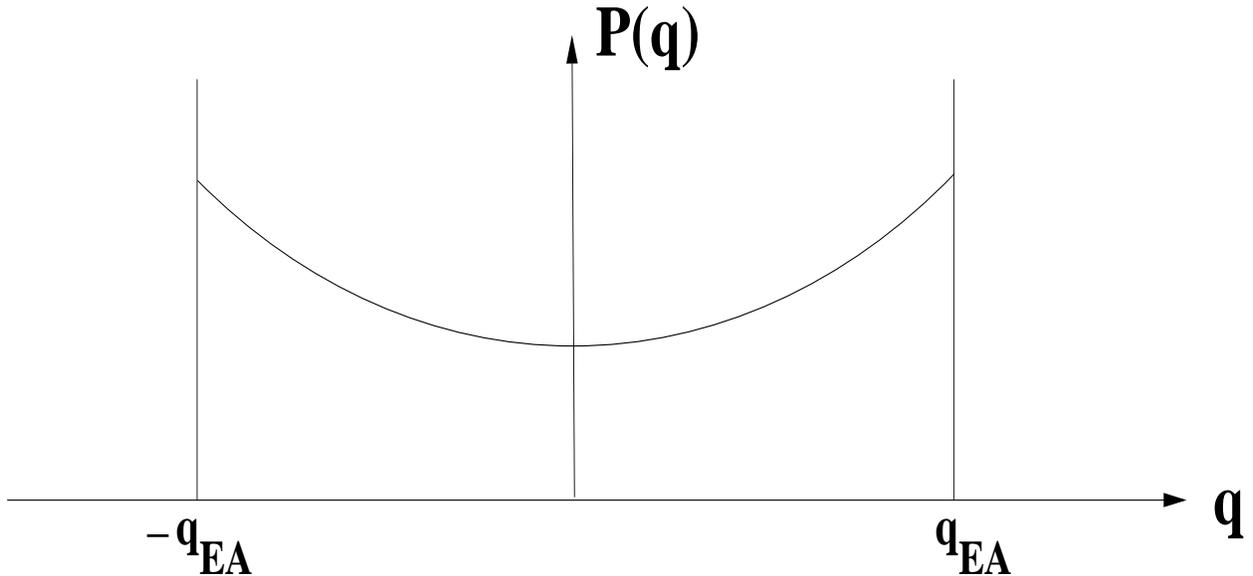,width=6.5in,height=3.0in}}
   \renewcommand{\baselinestretch}{1.0} 
   \small
   \caption{Sketch of the averaged overlap distribution function $P(q)$ for
   the SK model below $T_c$.  The spikes at $\pm q_{EA}$ are
   $\delta$-functions.}
   \end{figure}
   \renewcommand{\baselinestretch}{1.25}
   \normalsize

   Together $P_{{\cal J}}(q)$ and $P(q)$ can be thought of as describing the
   nature of ordering in the low-temperature phase of the SK model.  Instead
   of $P(q)$, one can study the function
   \begin{equation}
   \label{eq:xq}
   x(q)=\int^q P(q')dq'
   \end{equation}
   where $x(q)$ is essentially the fraction of states with overlap smaller
   than $q$.  This function, or more commonly its (monotonic) inverse $q(x)$,
   is also commonly referred to as the Parisi order parameter.  We will focus
   on $P(q)$ here.

   The sample-to-sample variation, even in the thermodynamic limit, implied by
   the non-self-averaging of the overlap distribution function, may seem
   somewhat disturbing at first, since it violates our thermodynamic intuition
   of a ``typical'' sample.  However, it should be remembered that $P_{\cal
   J}(q)$ is not directly measurable in the laboratory (though possibly
   information about it could be obtained indirectly).  Measurable
   thermodynamic quantities, such as free energy, magnetization, and so on,
   remain self-averaging.

   Of course, $P(q)$ {\it is\/} measurable through numerical simulations (see,
   for example, \cite{Yo83}).  Problems in the numerical determination of
   $P(q)$ for short-ranged models will be discussed in
   Sec.~\ref{subsec:problem}.

   \subsection{Ultrametricity}
   \label{subsec:ultra}

The discussion in the preceding subsection concerned random choices of
pairs of states taken from the Gibbs distribution $\rho_{{\cal J},N}(T)$.
A striking prediction of the Parisi solution concerns the expected outcome
when one independently chooses {\it triples\/} of states from $\rho_{{\cal
J},N}(T)$.  The disorder-averaged overlap probability distribution for
triples of states, $P(q_1,q_2,q_3)dq_1dq_2dq_3$, gives the probability that
the spin overlap between one of the three pairs of states lies in
$[q_1,q_1+dq_1]$, the second pair overlap in $[q_2,q_2+dq_2]$, and the
third in $[q_3,q_3+dq_3]$.  An explicit calculation of $P(q_1,q_2,q_3)$
using the Parisi RSB {\it ansatz\/} yields \cite{MPSTV84a,MPSTV84b}
\begin{eqnarray} 
\label{eq:ultra} P(q_1,q_2,q_3)&=&{1\over
2}P(q_1)x(q_1)\delta(q_1-q_2)\delta(q_1-q_3)\nonumber\\ &+&{1\over
2}P(q_1)P(q_2)\theta(q_1-q_2)\delta(q_2-q_3)\nonumber\\ &+&{1\over
2}P(q_2)P(q_3)\theta(q_2-q_3)\delta(q_3-q_1)\nonumber\\ &+&{1\over
2}P(q_3)P(q_1)\theta(q_3-q_1)\delta(q_1-q_2)\, .  
\end{eqnarray} 
That is,
if one chooses three states independently from $\rho_{{\cal J},N}(T)$, then
as $N\to\infty$ there is a probability $1/4$ that all overlaps are equal,
probability $1/4$ that $q_1=q_2<q_3$, probability $1/4$ that $q_2=q_3<q_1$,
and probability $1/4$ that $q_3=q_1<q_2$.  That is, if
$d_{\alpha\beta}=q_{EA}-q_{\alpha\beta}$ is defined as the ``distance''
between states $\alpha$ and $\beta$, then the distances among any three
states chosen randomly from $\rho$ form the sides of an equilateral
(probability $1/4$) or acute isosceles (probability $3/4$) triangle.  A
distance metric with these properties is called an ``ultrametric''; a
detailed discussion is given in \cite{RTV86}.  So there are strong
correlations among the states in the SK model, corresponding to a tree-like
or hierarchical structure among their overlaps \cite{vEHM92}.  This is illustrated in
Fig.~7.

   \begin{figure}
   \label{fig:ultra}
   \centerline{\epsfig{file=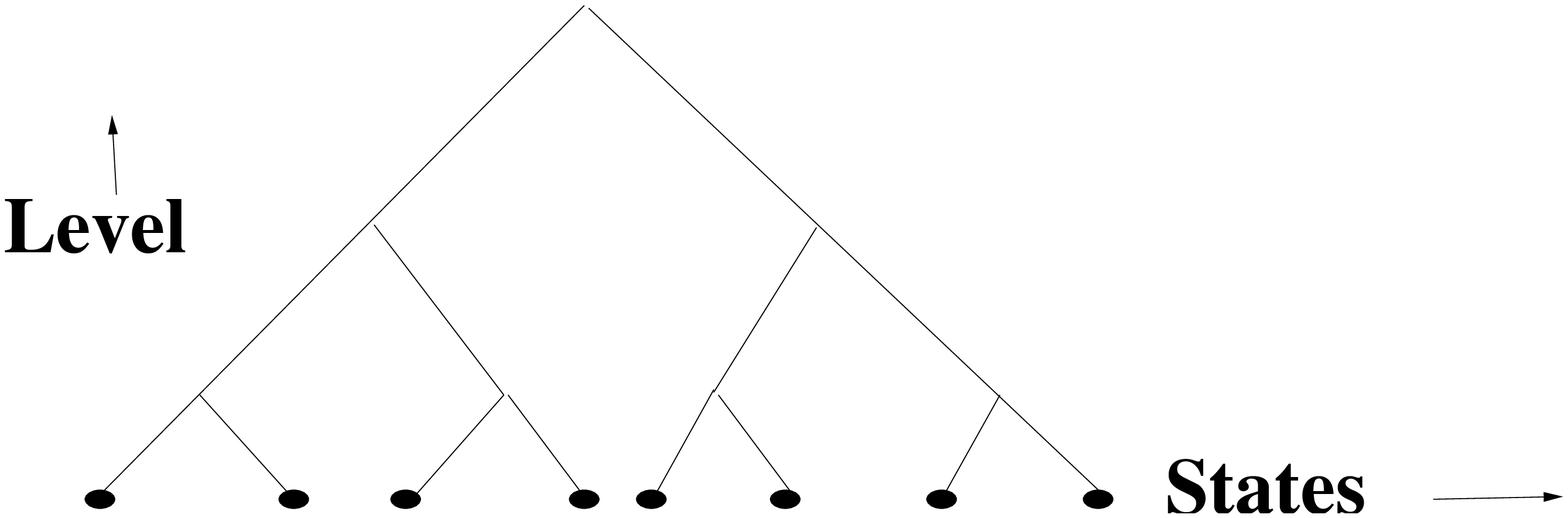,width=6.5in,height=3.0in}}
   \renewcommand{\baselinestretch}{1.0} 
   \small
   \caption{Rough sketch of the hierarchical structure of distances between states.
   A faithful representation of the actual tree corresponding to the Parisi
   solution would contain a continuous sequence of branchings.}
   \end{figure}
   \renewcommand{\baselinestretch}{1.25}
   \normalsize

   \medskip

   We have seen that the novel nature of the broken symmetry and ordering
   implied by the Parisi solution of the SK model is significantly different
   from anything observed in the low temperature phases of more familiar,
   homogeneous condensed matter systems.  It is therefore not surprising that
   the Parisi solution, once it was understood, generated considerable
   excitement.  Because mean-field theory has almost always provided the
   correct physics of the broken symmetry, ordering, and low temperature
   properties of more realistic models (typically providing poor quantitative
   results only close to $T_c$), it was natural to suppose that the RSB
   mean-field picture should similarly provide a correct description of the EA
   and other short-ranged spin glass models.  In the next few sections we will
   consider this issue.

   \section{Detection of Many States in Spin Glasses}
   \label{sec:csd}

   In describing the order parameters and broken symmetries of the SK spin
   glass phase, our starting point was the presumed existence of ``infinitely
   many pure thermodynamic states'' \cite{P83}.  However, in
   Sec.~\ref{subsec:pure} we cautioned about potential difficulties in
   applying these notions to the SK model.  Are the objections raised there
   merely pedantic, or do they carry substantial physical significance?  If
   the latter, how do we understand the physical meaning of the Parisi
   solution, and how can we apply it to the EA and other short-ranged models,
   where these concepts are better-defined?

   A second difficulty, also alluded to in Sec.~\ref{subsec:pure}, is the
   question of how one could actually construct a pure state in either the SK
   {\it or\/} the EA model.  In homogeneous systems like the uniform
   ferromagnet, this presents no difficulties: to select out the positively
   magnetized pure state, for example, one could either introduce a bulk term
   (a positive, homogeneous external field) whose magnitude goes to zero, or a
   surface term (fixed boundary condition consisting of all plus spins), with
   volume going to infinity.  However, even in the EA model, where the
   existence of well-defined pure states can be proved, there is no known
   ${\cal J}$-independent general procedure for constructing a single pure
   state if many are present.  This difficulty has also been posed in the
   context of broken ergodicity \cite{Pal82,PS85}.

   In approaching these questions, it is worthwhile to ask: how can one even
   know whether there {\it are\/} multiple pure state pairs?  Just as, in
   Sec.~\ref{subsec:overlaps}, one foregoes examination of individual pure
   states in the SK model for statistical information regarding their
   overlaps, we will similarly abandon the idea of attempting to construct
   individual pure states for a broader study of how the properties of the
   system are affected if many pure states exist.  To accomplish this we need
   to ask whether one can construct a simple and unambiguous procedure both
   for detecting the existence of multiple states, and for studying their
   properties.  In doing so, we will arrive at a deeper understanding of the
   meaning and consequences of the existence of many states in disordered
   materials; and we will simultaneously achieve a clearer, and deeper,
   understanding of the meaning and significance of the Parisi solution for
   the SK model.  The first step is showing the connection between the
   existence of multiple states and the presence of what we have called {\it
   chaotic size dependence\/} \cite{NS92}.

   \subsection{Chaotic Size Dependence in the SK model}
   \label{subsec:csdsk}

   \subsubsection{States}
   \label{subsubsec:states}

   We noted in Sec.~\ref{subsec:pure} that a thermodynamic state, pure or
   mixed, can be constructed as the infinite-volume limit of a sequence of
   finite-volume Gibbs distributions.  In the Curie-Weiss model, and again
   ignoring various technical complications in definitions of pure states, one
   can study the effects of adding a spin to an $N$-spin system, which
   simultaneously requires the introduction of $N$ new couplings.  (In spin
   glasses, this approach has been termed the ``cavity method'' \cite{MPV87}.)
   It is clear that addition of a single spin will not substantially alter a
   fixed correlation function, whether single- or multi-spin.  One can
   therefore sensibly conceive of a limiting thermodynamic state (in this case
   a mixture of positive and negative uniformly magnetized states, each with
   probability $1/2$).

   In the SK model, though, the situation is radically different.  As
   $N\to\infty$, any specified correlation function should not depend
   on any finite set of couplings, because the overall magnitude of
   any such finite set scales to zero and each of the new couplings
   accompanying each additional spin is chosen independently of the
   previous ones.  What this suggests is that (below $T_c$) any
   specified correlation function will not settle down to a limit as
   $N\to\infty$; equivalently, the cavity method does not result in a
   limiting thermodynamic state.  This heuristic argument introduces
   us to the pervasive presence of {\it chaotic size dependence\/} in
   spin glasses. For recent results on this issue, see~\cite{NS03}.

   If such a straightforward procedure does not result in a thermodynamic
   state, then does the notion of many pure states in the SK model make sense?
   In one way, at least, the answer is yes, even though the meaning of an
   individual pure state is still imprecise.  In an operational sense, one
   could in principle keep a record of the values of a finite set of even
   correlation functions (with, say, free boundary conditions) as $N$ grows
   (at fixed $T$).  If these values approach a limit, one can with some
   justification describe the low-temperature phase as consisting of a single
   pair of states (because each finite-spin state would simply be an equal
   mixture of the two).  If, however, the correlation functions persist in
   changing their values as $N$ grows, then one can infer that there must
   exist many such pairs --- presumably infinitely many, if there is no
   repetitive pattern.

   Now it is true, by compactness arguments, that in the latter case there
   must exist (infinitely many) subsequences that result in distinct limiting
   states.  But it is also almost certainly true that there is no {\it
   measurable\/} way of doing this; i.e., no procedure for selecting
   subsequences that is independent of a specific coupling realization ${\cal
   J}$.  The consequence is that there is {\it no finite procedure\/} for
   selecting convergent subsequences, and therefore generating thermodynamic
   states.

   However, while generating individual states seems to be problematic, it is
   nonetheless possible to devise simple procedures for measuring {\it
   statistical\/} properties of such states.  So the focus should not be on
   individual states, as in the conventional statistical mechanics of ordered
   systems, but instead on some larger construct that measures such
   statistical properties.  This will be further discussed in
   Sec.~\ref{sec:metastates}.

   \subsubsection{Overlaps}
   \label{subsubsec:over}

   States represent local quantities (i.e., correlation functions), and so
   depend on details of the coupling realization ${\cal J}$.  {\it Global\/}
   quantities, on the other hand, such as free energy per spin, should depend
   only on general parameters of the coupling {\it distribution} (such as its
   mean and variance), and therefore should have (the same) thermodynamic
   limit for almost every ${\cal J}$ \cite{ALR87,PS91,GT02}.  (From here on,
   we use the term ``almost every'', abbreviated ``a.e.'', in the strict
   probabilistic sense: that is, a result holding for ``for a.e. ${\cal J}$''
   means it holds for every coupling realization except for a set of measure
   zero in the space of all coupling realizations).

   What about a quantity like $P_{{\cal J},N}(q)$?  The fact that it's
   non-self-averaging already indicates that, although also a global measure,
   it may not have an $N\to\infty$ limit for a.e.~${\cal J}$ (using any
   coupling-independent procedure, i.e., one that is both measurable and
   finite).  This intuition was made rigorous in \cite{NS92}, as the following
   result:

   \medskip

   {\bf Theorem 1 \cite{NS92}.}  If $P_{{\cal J},N}(q)$ has a limit $P_{\cal
   J}(q)$ for a.e.~${\cal J}$, then it is self-averaged; i.e., $P_{\cal J}(q)$
   is independent of ${\cal J}$ for a.e.~${\cal J}$.

   \medskip

   {\bf Proof.} To prove this, we study $Y_{{\cal J},N}(t)$, the Laplace
   transform of $P_{{\cal J},N}(q)$:
   \begin{equation}
   \label{eq:y}
   Y_{{\cal J},N}(t)\equiv \int dq\ P_{{\cal J},N}(q)\ e^{tq}\, .
   \end{equation}
   Consider two coupling realizations, ${\cal J}$ and ${\cal J}'$, that differ
   in only finitely many couplings.  Then because the coupling magnitudes in
   the SK model scale as $N^{-1/2}$, it follows that
   \begin{equation}
   \label{eq:ratio}
   Y_{{\cal J}',N}(t)/Y_{{\cal J},N}(t)=e^{O(N^{-1/2})}\, 
   \end{equation}
   so if $Y_{\cal J}(t)=\lim_{N\to\infty}Y_{{\cal J},N}(t)$
   exists for a.e.~${\cal J}$, $Y_{\cal J'}(t)=Y_{\cal J}(t)$
   for any ${\cal J}$ and ${\cal J}'$ differing by a finite number of
   couplings.  But by the Kolmogorov zero-one law \cite{Fel71}, it follows
   that for each fixed $t$, $Y_{\cal J}(t)$ is constant for a.e.~${\cal
   J}$, which in turns implies that, for any $q$, $P_{\cal J}(q)$ is
   independent of ${\cal J}$ for a.e.~${\cal J}$. $\diamond$

   \medskip

   Here chaotic size dependence seems to follow from the scaling of the
   coupling magnitudes to zero as $N\to\infty$, with the result that no local
   state properties depend on a finite set of couplings.  While this is part
   of the story (and in fact, an important part), it is not the entire story.
   In short-ranged models like the EA spin glass, coupling magnitudes are
   fixed independently of volume, but as we shall see in the next section,
   chaotic size dependence is present there also, although certain differences
   now appear between the infinite- and finite-ranged models.

   \subsection{Chaotic Size Dependence in the EA model}
   \label{subsec:csdea}

   \subsubsection{``Observability'' of States}
   \label{subsubsec:observ}

   We already noted in the previous section that one important difference
   between the EA and the SK models is that coupling magnitudes scale to zero
   in the latter but not the former.  While this is also true in
   infinite-ranged models in homogeneous systems, the {\it randomness\/} in
   the couplings in the disordered case results in significant differences
   between short-ranged and infinite-ranged models.

   As before, consider a volume $\Lambda_L$ in the EA model.  Boundary
   conditions, such as free, periodic, antiperiodic, or fixed, now need to be
   specified.  (In some rough sense, one can think of the SK model as having
   free ``boundary'' conditions.)  Recall that in the homogeneous Ising
   ferromagnet below $T_c$, different pure states can be generated by using
   sequences of plus {\it vs.\/}~minus boundary conditions.  So one test of multiple
   states is the {\it sensitivity\/} of correlation functions, deep in the
   interior, upon change of boundary conditions far away.

   If there is only a single pure state (e.g., a paramagnet), then any
   sequence of boundary conditions results in that single limiting state, and
   there is no chaotic size dependence.  If broken spin-flip symmetry exists,
   i.e., there is a phase with $q_{EA}>0$, then there must be different
   sequences of boundary conditions leading to different limiting states.  But
   how can we use this fact to determine whether there is only a single pair of
   pure states, as in the droplet/scaling picture, or infinitely many pairs,
   as would be required, for example, by an RSB-like picture?

   We start by defining an {\it observable\/} state.  We do not expect
   observable physical properties to depend on the microscopic details
   of the couplings, but instead on macroscopic properties.  For
   example, in a dilute magnetic alloy, one needs to study the system
   without knowing the microscopic locations of all the magnetic
   impurity atoms.  As a consequence, measurements on a spin glass are
   necessarily made in a manner independent of the microscopic
   coupling realization corresponding to that particular sample.

   This suggests that physical properties should be associated with states
   that are mathematically constructed in a {\it coupling-independent\/}
   manner; in particular, by using boundary conditions that do not depend
   on the couplings \cite{vEFr85,GNS93}.  We are therefore led to the
   following:

   {\bf Working Definition:} An {\it observable state\/} is a thermodynamic state,
   pure or mixed, whose existence is manifested through some sequence of
   coupling-{\it independent\/} boundary conditions.

   This working definition is not mathematically precise, but will be made so in
   Sec.~\ref{subsubsec:physobserv}.  The definition suggests the possible
   existence of other, ``invisible'', states; these will be briefly discussed
   in Sec.~\ref{subsec:invisible}.

   Clearly, the many states appearing in the Parisi solution of the SK model
   are observable.  But this then leads to another question: instead of
   bothering with boundary conditions, why not use the overlap measure
   $P_{\cal J}(q)$ to detect many states in the EA model, just as it was
   successfully used in the SK model?  As we will see in
   Sec.~\ref{subsec:problem}, $P_{\cal J}(q)$ turns out 
   in general to be an unreliable
   indicator of the existence of many states in short-ranged models (whether
   homogeneous or disordered); it can give the appearance of many
   states where there is only a single pair, and the appearance of only a
   single pair --- indeed, sometimes even only a single state --- when there
   are infinitely many.

   So how can one distinguish between the presence of infinitely many pure
   states {\it vs.\/}~only a single pair for the EA model?  In the next two subsections,
   we will show that the answer lies in studying {\it spin-flip symmetric\/}
   boundary conditions.

   \subsubsection{Sensitivity to Boundary Conditions and ``Windows''}
   \label{subsubsec:windows}

   The definition of observable states given in the previous section still
   seems somewhat impractical, or at least not ``finite'', since it involves
   the existence of infinite-volume limits.  However, its consequences are
   indeed practical {\it and\/} finite, in particular for numerical
   experiments.  For example, one implication is that the presence of multiple
   obervable states manifests itself as a sensitivity of correlation functions
   deep in the interior of a finite volume $\Lambda_L$ to changes in the
   boundary conditions on $\partial\Lambda_L$.

   In order to make this idea precise, consider a fixed volume $\Lambda_{L_0}$
   within which correlation functions are studied, centered inside a much
   larger volume $\Lambda_L$.  We will call $\Lambda_{L_0}$ a ``window''
   \cite{NS98}.

   Suppose now that there exists only a single pure state at some temperature
   $T$; for sufficiently high $T$ in either the EA or SK models, this would be
   the paramagnet.  As noted in the previous section, any infinite sequence of
   boundary conditions will result in this single limiting pure state.  (In
   the above sentence, and throughout the paper, ``infinite sequence of
   boundary conditions'' is shorthand for an infinite sequence of
   finite-volume Gibbs states, as in Eq.~(\ref{eq:finite}), generated by a
   sequence of volumes with the specified boundary conditions.)  As a
   consequence, if one looks at a fixed window $\Lambda_{L_0}$ inside a much
   larger volume $\Lambda_L$, any change in the boundary conditions on
   $\partial\Lambda_L$ will leave the correlation functions inside the window
   largely unaffected (with any small effects vanishing as the boundaries
   recede to infinity.)

   Suppose now that there are two pure states, say the positive and negative
   magnetization states in the Ising ferromagnet.  If one switches from the
   fixed boundary condition on $\partial\Lambda_L$
   with all boundary spins $+1$, to the fixed
   boundary condition with all boundary spins $-1$, correlation functions {\it
   everywhere\/} inside the volume will change (corresponding to a change from
   the positively magnetized to the negatively magnetized state), no matter
   how large $L$ becomes.

   So the presence of multiple Gibbs states results in sensitivity to
   boundary conditions.  Demonstration of such sensitivity in spin glasses at
   sufficiently low but nonzero temperature in a given dimension would already
   be sufficient to answer Open Question 1.  More is needed, however, if one
   wants to address the issue of whether there is a single pair of pure states
   or infinitely many.

   \subsubsection{Domain Walls and Free Energies}
   \label{subsubsec:free}

   As noted in Sec.~\ref{subsec:overlaps}, an infinite sequence of periodic
   boundary conditions in the (uniform or random) Ising ferromagnet, above the
   lower critical dimension and below $T_c$, does lead to a limiting mixed
   state $\rho_{\rm per}={1\over 2}\rho_+ +{1\over 2}\rho_-$, where $\rho_+$
   is the positively magnetized and $\rho_-$ the negatively magnetized state.
   If these are the {\it only\/} pure states, then any sequence of
   antiperiodic boundary conditions will also have a limit and yield the {\it
   same\/} mixed state.  Of course, there will be a relative domain wall
   between periodic and antiperiodic b.c.~states in a fixed volume, but as the
   volume size increases, the domain wall eventually moves outside of any
   fixed window.  In such a scenario we say the domain wall has ``deflected to
   infinity''.

   One can prove a similar result in the EA Ising spin glass with periodic,
   antiperiodic, free, or other ``symmetric'' boundary conditions.  By
   symmetric b.c.~we mean one for which the resulting Gibbs state $\rho_L$ in
   any $\Lambda_L$ is spin-flip-invariant; that is, all odd correlations
   vanish.  In a spin glass, if there is only a single pair of pure states
   $\rho$ and $\overline{\rho}$ that transform into each other under a global
   spin flip (as in the droplet/scaling picture), then any sequence of
   symmetric boundary conditions has a limiting thermodynamic state
   \cite{NS92}, which is simply the mixed state $\rho_{\rm sym}={1\over
   2}\rho + {1\over 2}\overline{\rho}$.  As a consequence, switching from
   periodic to antiperiodic b.c.'s in a large volume leaves the Gibbs
   state unaffected inside a window deep in the interior.

   What happens if there are many pure state pairs, as would be the case in a
   mean-field-like picture?  In this case, two arbitrarily chosen pure states
   (not from the same pair) would have relative domain walls that do {\it
   not\/} deflect to infinity.  If the free energy cost of such domain walls
   does not exceed the free energy difference in a typical $\Lambda_L$
   incurred by switching from, say, periodic to antiperiodic boundary
   conditions, then these different pairs of states should be observable under
   such a switch.  In other words, if many pure state pairs exist, and their
   relative free energy differences are not too large, then switching from
   periodic to antiperiodic b.c.'s in a typical $\Lambda_L$ should change the
   Gibbs state deep in the interior.

   How large is ``not too large''?  A trivial upper bound is a domain wall
   whose free energy scales as $L^{d-1}$ or smaller, since a switch in
   boundary conditions obviously cannot result in a greater free energy
   change.  But in the EA model, a much better bound can be found rigorously.
   At fixed temperature, if one switches from periodic to antiperiodic b.c.'s
   then in $\Lambda_L$, the root mean square free energy difference is
   bounded from above by $O(L^{(d-1)/2})$, i.e., the square root of the surface
   area \cite{NS92,AF91}.  This rigorous result, which we present here without
   proof, and which applies to other ``flip-related'' \cite{NS92}
   coupling-independent b.c.'s, supports Anderson's \cite{And78} intuitive
   notion of frustration described in Sec.~\ref{subsec:frustration}.

   This suggests the following picture: in dimensions with broken
   spin-flip symmetry, there exists an exponent $\theta(d)$ with
   $0<\theta(d)\le (d-1)/2$ \cite{FH86,FH88} that governs free energy
   fluctuations upon a switch from periodic to antiperiodic b.c.'s.
   If many pure state pairs exist, and their lowest-lying
   relative domain wall energies scale as $L^\theta$ or less, then the
   Gibbs state $\rho_L$ in a typical $\Lambda_L$ will be sensitive
   throughout the entire volume to changes in coupling-independent
   b.c.'s, irrespective of the size of $L$. In other words, these many
   pure state pairs are observable in the sense described in
   Sec.~\ref{subsubsec:observ}.
     
     What is expected in a mean-field-like scenario for the EA model?  If the
     many pure state pairs there are not observable, a case could be made that
     they are unphysical and not of great interest.  However, because they {\it
     are\/} seen in the SK model under what can roughly be thought of as
     (coupling-independent) free boundary conditions, one would expect the
     analogous states --- if they exist --- in the EA model to be observable
     also.  In fact, RSB theory predicts that domain walls between such states
     are both ``space-filling'' (to be discussed in Sec.~\ref{sec:interfaces})
     and with energies of $O(1)$ {\it independent\/} of lengthscale
     \cite{MP00a,MP00b,PY00,KM00}.  They therefore must manifest themselves
     under a change from, say, periodic to antiperiodic boundary conditions
     \cite{MP00a} in a typical large $\Lambda_L$.

     \subsubsection{Many States and Chaotic Size Dependence}
     \label{subsubsec:manystates}

     The discussion so far indicates that the existence of many pairs of
     ``SK-like'' pure states in the EA model leads to chaotic size dependence:
     if one takes a sequence of volumes all with, e.g., periodic boundary
     conditions (with the volumes chosen deterministically, i.e., independently
     of the couplings) then there will almost surely be no limiting
     thermodynamic state.  That is, a typical $n$-spin correlation function,
     computed in each $\Lambda_L$ using the corresponding finite-volume Gibbs
     state, will change continually as $L \to \infty$, never settling down to a limit.

     This leads to an interesting alternative formulation of the problem of
     existence of many pure state pairs in the EA model.  Consider any arbitrary
     deterministic sequence of volumes with symmetric (periodic, antiperiodic,
     free, or other) b.c.'s.  {\it If there exists only a single pair\/}
     $\alpha,\overline{\alpha}$ {\it of (observable) states, as in the
     droplet/scaling model, then for fixed\/} ${\cal J}$ {\it such a sequence
     will (with probability one over the\/} ${\cal J}${\it 's) possess a limit:
     the thermodynamic state\/} $\rho={1\over 2}\alpha + {1\over
     2}\overline{\alpha}$.  {\it If instead there are two or more (observable)
     pure state pairs, then there will be chaotic size dependence: with
     probability one, such a sequence will not converge to a limiting
     thermodynamic state.\/}

     The precise statement underlying these conclusions is the following:

     {\bf Theorem 2 \cite{NS92}.} Given a symmetric coupling distribution,
     suppose that a sequence of coupling-independent boundary conditions results
     in a limiting thermodynamic state $\rho$.  If in every volume in the
     sequence (or an arbitrary subset chosen independently of the couplings),
     the original boundary condition is replaced by one that is flip-related,
     then (1) the new sequence will also have a limit, which (2) will be the same
     thermodynamic state $\rho$.

     {\it Remark.\/} By ``flip-related'' b.c.'s we mean that for each $L$, there
     is some $B_L\subset\partial\Lambda_L$ whose flip transforms one b.c.~into
     the other.  So, for example, periodic and antiperiodic b.c.'s are
     flip-related; so are two different fixed b.c.'s.  However, periodic and
     fixed b.c.'s are not flip-related.

     The proof of the second conclusion is fairly straightforward, and will be
     useful later, so we present it here.  For a proof of the first conclusion,
     see \cite{NS92}.

     {\bf Proof of Part 2.}  Consider the correlation function of the spins at
     the $m$ sites $i_1,\ldots,i_m$.  For each $L$, call this $m$-point
     correlation function $X_L^1$ for b.c.~1 and $X_L^2$ for b.c.~2, which is
     flip-related to b.c.~1.  $X_L^1$ and $X_L^2$ are bounded between $-1$ and
     $+1$ for all $L$ and all ${\cal J}$.  The limits
     $X^1=\lim_{L\to\infty}X_L^1$ and $X^2=\lim_{L\to\infty}X_L^2$ exist by
     assumption, and are functions of ${\cal J}$.  Let $X({\cal J})\equiv
     X^1({\cal J})-X^2({\cal J})$.

     Consider now the conditional expectation $E_r[\cdot]$, defined as the
     expectation of a quantity after averaging over all couplings {\it
     outside\/} $\Lambda_r$; so $E_r[\cdot]$ depends on the couplings only inside
     $\Lambda_r$.  Then
     \begin{equation}
     \label{eq:Er}
     E_r[X]=\lim_{L\to\infty}E_r[X_L]=\lim_{L\to\infty}\left(E_r[X_L^1]-E_r[X_L^2]\right)
     \end{equation}
     exists.  Because $r$ and $m$ are fixed, and $L\to\infty$, eventually $L>r$
     and $\{i_1,\ldots,i_m\subset\Lambda_L\}$ for any $r$ and $m$.  But because
     $E_r[\cdot]$ averages over the boundary bonds of the cube, and the b.c.'s 1
     and 2 are flip-related, it is easy to see that 
     $E_r[X_L^1] = E_r[X_L^2]$ for large $L$ and hence that $E_r[X]=0$ for every $r$.
     If a bounded random variable $X$ has zero conditional expectation for every
     $r$, then $X=0$ (for a.e.~${\cal J}$).  The claimed result follows, because the
     same argument holds for all correlation functions.  $\diamond$

     Chaotic size dependence in the presence of many states is intuitively
     plausible.  Suppose that there exists infinitely many pure state pairs.
     For a given $\Lambda_L$, some subset of those states will have larger
     weights than others; in a rough sense, they will be those which best
     ``match'' the boundary condition for that $L$.  
     As $L$ varies, the selected states with
     large weights should vary in some unpredictable fashion.  The theorem
     simply proves this plausible scenario.

     There is a numerical consequence of this observation.  Just as the presence
     of many states is in principle detectable numerically by looking at
     sensitivity of the state in the deep interior to changes in the boundary
     condition in a {\it fixed\/} volume (Sec.~\ref{subsubsec:free}), one can
     also look numerically for chaotic size dependence; that is, study a given
     set of correlation functions as the volume size changes for a {\it fixed\/}
     b.c.~(such as periodic).

     It may now seem that chaotic size dependence adds a layer of complexity to
     the study of thermodynamic states in spin glasses.  How can one even talk
     about pure states when there now seems to be no measurable way to construct
     them (if there are infinitely many)?  A new thermodynamic tool seems to be
     needed; such a tool will be presented in the next section.

     \section{Metastates}
     \label{sec:metastates}

     \subsection{Motivation and Mathematical Construction}
     \label{sub:math}

     If there exist many observable pure states, a sequence of
     coupling-independent b.c.'s will generally {\it not\/} converge to a
     limiting thermodynamic state: there is chaotic size dependence (hereafter
     denoted CSD).  That is, a typical correlation
     $\langle\sigma_{i_1}\ldots\sigma_{i_n}\rangle$, computed in $\Lambda_L$
     from the finite-volume Gibbs state, will not have a single limit as
     $L\to\infty$ but rather many different limits along different subsequences
     of $L$'s (chosen in a coupling-dependent manner).

     Such behavior in $L$ is analogous to chaotic behavior in time $t$
     along the orbit of a dynamical system.  Of course, in each case the
     behavior is deterministic but effectively unpredictable, and appears
     to be a random sampling from some distribution $\kappa$ on the space
     of states.  In the case of dynamical systems, one can in principle
     reconstruct $\kappa$ by keeping a record of the proportion of time the
     particle spends in each coarse-grained region of state space.
     Similarly, one can prove \cite{NSBerlin,NS96b} that for inhomogeneous
     systems like spin glasses, a similar distribution exists: roughly
     speaking, the fraction of $\Lambda_L$'s in which a given thermodynamic
     state $\Gamma$ appears converges, even in the presence of CSD.  By
     saying that a thermodynamic state $\Gamma$ (which is an
     infinite-volume quantity) ``appears'' within a finite volume
     $\Lambda_L$, we mean the following: within a window deep inside the
     volume, all correlation functions computed using the finite-volume
     Gibbs state $\rho_L$ are the same as those computed using $\Gamma$
     (with negligibly small deviations).  The state $\Gamma$ can be either
     pure or mixed, depending on the boundary conditions.

     Mathematically, a metastate $\kappa$ is a probability measure on the space
     of all (fixed ${\cal J}$) thermodynamic states.  Of course, the metastate
     depends on the boundary conditions used: we will refer to the metastate
     constructed from a deterministic sequence of volumes, all with periodic
     boundary conditions, as the ``periodic b.c.~metastate'', and similarly for
     the antiperiodic b.c.~metastate, the free b.c.~metastate, and so on.  One
     can also construct metastates in which the b.c.'s vary with $L$.

     A simple  empirical construction of $\kappa$ would be as follows: consider a
     ``microcanonical'' ensemble $\kappa_L$, in which each of the finite volume
     Gibbs states $\rho^{(1)}, \ldots ,\rho^{(L)}$ in volumes
     $\Lambda_1,\ldots,\Lambda_L$ has weight $L^{-1}$.  Then
     $\kappa=\lim_{L\to\infty}\kappa_L$.  The meaning of the limit is that for
     every well-behaved function $g(\cdot)$ on states,
     \begin{equation}
     \label{eq:macro1}
     \lim_{L\to\infty}L^{-1}\sum_{{\ell}=1}^{L} g(\rho^{({\ell})}) =
     \{g(\Gamma)\}_\kappa\quad ,
     \end{equation}
     where the bracket $\{ . \}_\kappa$ denotes the average over $\kappa$.

     There is an alternative (and earlier) construction due to Aizenman and Wehr
     \cite{AW90}.  In this construction, one replaces the microcanonical
     ensemble $\kappa_L$ by the ensemble of states obtained by varying the
     couplings {\it outside\/} $\Lambda_L$. The limit here means that for every
     well-behaved function $F$ of finitely many couplings and finitely many
     correlations,
     \begin{equation}
     \label{eq:macro2}
     \lim_{L'\to\infty}[F({\cal J},\rho^{(L')})]_{av} = 
     \left[\{F({\cal J},\Gamma)\}_{\kappa({\cal J})}\right]_{av} \quad .
     \end{equation}
     Here, $[ . ]_{av}$ denotes the average over the quenched coupling
     distribution.  In fact it has not yet been proved that these relatively
     simple limits, using all $\ell$'s and $L$'s , are valid.
     However, it can be proved \cite{NSBerlin} that there exist
     deterministic (i.e., ${\cal J}$-independent) subsequences of ${\ell}$'s and
     $L$'s for which limits such as in both Eq.~(\ref{eq:macro1}) and
     Eq.~(\ref{eq:macro2}) exist and yield the {\it same\/} $\kappa({\cal J})$.

     \subsection{Physical Meaning and Significance}
     \label{sub:phys}

     \subsubsection{Observable States and Thermodynamic Chaos}
     \label{subsubsec:physobserv}

     Like a thermodynamic state $\Gamma$, the metastate $\kappa$ is an
     infinite-volume probability measure.  But while $\Gamma$ is a measure on
     spin configurations, $\kappa$ is a measure on the {\it thermodynamic
     states\/} themselves.  That is, $\Gamma$ provides the probability that a
     given spin configuration appears inside a finite region, while the
     metastate $\kappa$ provides the probability that a given pure or mixed
     state appears inside a typical large volume $\Lambda_L$ with specified
     b.c.'s.  As such, the metastate contains far more information than any
     single thermodynamic state.

     So, instead of treating CSD as a problem and trying to do an ``end run''
     around it, introducing metastates allows us to exploit the vast amount of
     information contained within CSD; for fixed ${\cal J}$, a metastate allows
     us to analyze how the finite volume Gibbs states $\rho_L$ with given
     boundary conditions sample from the available set of thermodynamic states.
     Although the metastate concept is equally applicable to situations where
     CSD does {\it not\/} occur, it is most useful as a tool for analyzing
     ``thermodynamic chaos''.

     As always, we are interested in metastates constructed using
     coupling-independent boundary conditions.  This allows us to redefine 
     somewhat more
     precisely the notion of an observable state roughly defined in
     Sec.~\ref{subsubsec:observ}:

     {\bf Definition:} An {\it observable state\/} is a thermodynamic state,
     pure or mixed, that lies in the support of some coupling-independent
     metastate.

     \subsubsection{Finite vs.~Infinite Volumes}
     \label{subsubsec:finite}

     Another useful consequence of using metastates is that they enable us to
     relate the observed behavior of a system in large but finite volumes with
     the system's thermodynamic properties.  This relation is relatively
     straightforward for systems with a few pure states or for those whose
     states are related by well-understood symmetry transformations; but in the
     presence of many pure states not related by any obvious transformations,
     this relation may be subtle and complex.  Here the metastate approach may
     be not only useful but necessary.

     Occasionally a distinction is drawn between finite- and infinite-volume
     states (see, for example, \cite{MPR97}), where it is argued that the first
     is more physical and the second merely mathematical in nature.  While we
     will see below (see also \cite{NSBerlin,NS92,NS98,NS96b,NS97}) that the
     relation between the two may be more subtle than previously realized, we
     also argue that such a distinction is misleading.  Indeed, it should be
     clear from the discussion above that the metastate approach is specifically
     constructed to consider both finite and infinite volumes together and to
     unify the two cases.

     \section{Can a Mean-Field Scenario Hold in Short-Ranged Models?}
     \label{sec:mfscenario}

     We have now developed the tools we need to analyze whether the type of
     ordering present in the RSB solution of the SK model can hold in more
     realistic short-ranged models.  The two most-discussed scenarios have been
     the many-state mean field picture described above, and the two-state
     droplet/scaling picture introduced in Sec.~\ref{sec:open}.  We will see
     below that application of the RSB picture to short-ranged models is not at
     all straightforward, resulting in considerable confusion in the literature.
     Before we turn to that subject, however, we need to examine a preliminary
     question: at fixed dimension and temperature, why can't it be that the
     mean-field scenario holds for, say, half of all coupling realizations and
     droplet/scaling for the other half?

     \subsection{Translation-Ergodicity}
     \label{subsec:transerg}

     In fact, such a possibility {\it cannot\/} occur: any type of ordering,
     based on multiplicity of states, whether it's one of the above or something
     else entirely, {\it must occur either for every ${\cal J}$\/} (except for a
     set of measure zero) {\it or for none.\/} The proof of such a statement lies in
     a straightforward use of the ergodic theorem \cite{Nadkarni98}.  Because
     the ideas used here will be useful later on, we make a small detour to
     explain it more fully.

     As always, we assume a fixed coupling distribution $\nu({\cal J})$, in
     which the couplings are independent, identically distributed random
     variables.  At some fixed dimension $d$ and temperature $T$, let ${\cal
     N}(d,T,{\cal J})$ denote the number of pure states (one or two or
     $\ldots\infty$).  Then it can be proved that ${\cal N}(d,T,{\cal J})$ is
     constant almost surely; i.e., it is the same for a.e.~${\cal J}$, at fixed
     $T$ and $d$.  (Of course, ${\cal N}(d,T,{\cal J})$ can and --- if there is
     a low-temperature spin glass phase --- will have some dependence on both
     $d$ and $T$.  It also clearly can depend on $\nu$, although that dependence
     is not explicitly indicated.)  In the language of spin glass theory, ${\cal
     N}$ is a self-averaged quantity.

     The proof depends on three ingredients: measurability of ${\cal N}$ as a
     function of ${\cal J}$, translation-invariance of ${\cal N}$ with respect
     to a uniform shift of the couplings, and translation-ergodicity of the
     underlying disorder distribution ${\nu}$.  A discussion of measurability
     (in the mathematical sense) would be somewhat technical and will be avoided
     here, and a precise definition will not be given; but, roughly speaking, it
     implies, for a function on random variables, that there is an explicit
     realization-independent procedure for constructing it.  A proof that ${\cal
     N}$ is measurable is given in \cite{NSBerlin}.

     The concepts of translation-invariance and translation-ergodicity are
     relatively straightforward, and we discuss them informally here.  Let $a$
     be any lattice translation; then ${\cal J}^a$ indicates the coupling
     realization with the locations of
     all couplings in ${\cal J}$ uniformly shifted by $a$.  A
     translation-invariant function $f$ on ${\cal J}$ is one where $f({\cal
     J}^a)=f({\cal J})$ and a translation-invariant distribution
     of ${\cal J}$'s is defined similarly.  
     Clearly, both ${\cal N}$ and the disorder distribution
     $\nu$ are translation-invariant.

     Translation-{\it ergodicity\/} of a probability measure, such as $\nu$, is
     analogous to the more familiar notion of time-ergodicity.  Consider again a
     function $g$ of ${\cal J}$, with ${\cal J}$ chosen from some distribution $\mu$;
     $g$ may or may not be translation-invariant.
     The {\it distribution\/}-average $E_\mu[g]$ of $g$ is what we normally
     refer to simply as the average; that is 
     \begin{equation}
     \label{eq:distav}
     E_\mu[g]=\int d{\cal J}\,\mu({\cal J})g({\cal J})\, .
     \end{equation}
     But a {\it translation\/}-average $E_t[g]$ can also be defined for any
     realization ${\cal J}$, as simply the spatial average of $g({\cal J})$ under all
     lattice translations ${\cal J}^a$.  Then the distribution $\mu$ is
     translation-ergodic if
     \begin{equation}
     \label{eq:transerg}
     E_\mu[g]=E_t[g]\ ,
     \end{equation}
     for a.e.~${\cal J}$.

     By the ergodic theorem, any (measurable) translation-invariant function 
     of ${\cal J}$ chosen from a translation-ergodic distribution is constant
     almost surely (that is, is the same for a.e.~realization of ${\cal J}$).

     Informally, this is easy to see: by definition, a translation-invariant
     function is constant with respect to any lattice translation of the
     realization on which it depends.  Suppose that $g({\cal J})$ is
     translation-invariant for a.e.~${\cal J}$, and suppose (for example) that
     it equals the constant $g_1$ for half of all realizations and $g_2$ for the
     other half, with $g_1\ne g_2$.  Then the distribution-average
     $E_\mu[g]=(1/2)(g_1+g_2)$, which is not equal to $g({\cal J})$ for {\it
     any\/} ${\cal J}$ (outside of a possible set of measure zero).  But this
     violates Eq.~(\ref{eq:transerg}), contradicting the supposition that $\mu$
     is a translation-ergodic distribution.

     Returning to the question of the possible variability of the number of pure
     states with ${\cal J}$, we recall that the couplings are independent,
     identically distributed random variables.  The distribution for such
     random variables is translation-ergodic \cite{Wie39}, and so $\nu({\cal
     J})$ is translation-ergodic.  So, because ${\cal N}$ is a
     translation-invariant function of ${\cal J}$, which is drawn from the
     translation-ergodic distribution $\nu({\cal J})$, it follows that ${\cal
     N}$ is the same for a.e.~${\cal J}$ (at fixed $T$ and $d$).  Of course,
     this argument doesn't tell us the {\it value\/} of ${\cal N}$ at a given
     $T$ and $d$, only that it's constant with respect to ${\cal J}$.

     This argument has been presented in some detail not because there is any
     controversy on this particular question -- there isn't -- but because
     similar arguments can be used to resolve issues that heretofore {\it had\/}
     been controversial.  We will turn to these issues in the following
     sections.  Before doing so, we note that these and similar results allow us
     to make statements like ``The number of pure states in a spin glass at low
     temperatures in three dimensions is $x$.''  What this statement really
     means is that, with probability one, for any particular realization of the
     couplings on a three-dimensional lattice, there are $x$ pure states at that
     temperature.  (Even a statement such as this, however, should specify
     whether one is talking about observable states only, but that will be
     assumed in what follows.  We conjecture, but have not proved, that the same
     number of pure states will be in the support of a.e.~coupling-independent
     metastate.)

     \subsection{The Standard SK Picture}
     \label{subsec:standard}

     We now turn to an examination of whether a mean-field-like picture can hold
     in short-ranged spin glasses like the EA model.  It may seem initially that
     the outlines of such a picture should be clear.  A typical description is
     given in \cite{FMPP98}:

     {\narrower\smallskip\noindent
     ``Hence the Gibbs equilibrium measure decomposes into a mixture of many
     pure states.  This phenomenon was first studied in detail in the mean field
     theory of spin-glasses, where it received the name of replica-symmetry
     breaking $\ldots$ But it can be defined in a straightforward way and easily
     extended to other systems, by considering an order parameter function, the
     overlap distribution function.  This function measures the probability that
     two configurations of the system, picked up independently with the Gibbs
     measure, lie at a given distance from each other $\ldots$ Replica-symmetry
     breaking is made manifest when this function is nontrivial.''\smallskip}

     But if there are many pure states, then there must be CSD; and if there is
     CSD, what does one mean by the equilibrium Gibbs measure?  This question
     was first addressed in \cite{NS96c}, where it was shown that the most
     natural and straightforward interpretation of statements like the one above
     --- what we have called the standard SK picture --- {\it cannot\/} hold in
     short-ranged models in any dimension and at any temperature.

     The difficult part in studying this problem is that of constructing
     limiting Gibbs states, given the presence of CSD when there are many
     states --- but the notion of the metastate now makes it easy (or at least
     easier).  In \cite{NS96c}, two constructions of overlap distributions were
     given, but we use metastates here to simplify the discussion.  Before
     proceeding, we will use the description above to construct the standard SK
     picture.

     The paragraph from \cite{FMPP98} quoted above, as applied to the EA model at fixed
     temperature $T$, requires a Gibbs equilibrium measure $\rho_{\cal
     J}(\sigma)$ which is decomposable into many pure states $\rho_{\cal
     J}^\alpha (\sigma)$:
     \begin{equation}
     \label{eq:sum}
     \rho_{\cal J}(\sigma)=\sum_\alpha W_{\cal J}^\alpha\rho_{\cal J}^\alpha (\sigma)\ .
     \end{equation}

     One then considers the overlap distribution function $P_{\cal J}(q)$,
     constructed as described above.  That is, one chooses $\sigma$ and
     $\sigma'$ from the product distribution $\rho_{\cal J}(\sigma)\rho_{\cal
     J}(\sigma')$, and then the overlap
     \begin{equation}
     \label{eq:overlap}
     Q=\lim_{L\to\infty}|\Lambda_L|^{-1}\sum_{x\in\Lambda_L}\sigma_x\sigma'_{x}
     \end{equation}
     has $P_{\cal J}$ as its probability distribution.  Here $|\Lambda_L|$ is the
     volume of the cube $\Lambda_L$.  

     Given Eq.~(\ref{eq:sum}), there is a nonzero probability that $\sigma$ and
     $\sigma'$ will be chosen from different pure states.  If $\sigma$ is drawn
     from $\rho_{\cal J}^\alpha$ and $\sigma'$ from $\rho_{\cal J}^\beta$, then
     the expression in Eq.~(\ref{eq:overlap}) equals its thermal mean,
     \begin{equation}
     \label{eq:qab}
     q_{\cal J}^{\alpha\beta}=\lim_{L\to\infty}|\Lambda_L|^{-1}
     \sum_{x\in\Lambda_L} \langle\sigma_x\rangle_\alpha
     \langle\sigma_x\rangle_\beta \quad .
     \end{equation}
     Thus $P_{\cal J}$ is given by
     \begin{equation}
     \label{eq:PJ(q)}
     P_{\cal J}(q)=\sum_{\alpha,\beta}W_{\cal J}^\alpha W_{\cal J}^\beta
     \delta(q-q_{\cal J}^{\alpha\beta})\quad .
     \end{equation}

     In the mean-field picture, the $W_{\cal J}^\alpha$'s and $q_{\cal
     J}^{\alpha\beta}$'s are non-self-averaging quantities, except for
     $\alpha=\beta$ or its global flip, where $q_{\cal J}^{\alpha\beta}=\pm
     q_{EA}$.  As in Sec.~\ref{subsec:nsa}, the average $P(q)$ of $P_{\cal
     J}(q)$ over the disorder distribution $\nu$ of the couplings is a mixture
     of two delta-function components at $\pm q_{EA}$ and a continuous part
     between them.

     The problem, as already noted, is constructing $\rho_{\cal J}(\sigma)$
     given the presence of CSD: simply taking a sequence of cubes with periodic
     b.c.'s, for example, won't work.  However, consider the periodic b.c.~{\it
     metastate\/} $\kappa^{\rm PBC}_{\cal J}$ (in fact, any coupling-independent
     metastate would do).  One can construct a state $\rho_{\cal
     J}(\sigma)$ which is the {\it average\/} over the metastate: 
     \begin{equation}
     \label{eq:rhoav}
     \rho_{\cal J}(\sigma)=\int\ \Gamma(\sigma)\kappa_{\cal J}(\Gamma)\ d\Gamma\quad .
     \end{equation} 
     One can also think
     of this $\rho_{\cal J}$ as the average thermodynamic state,
     $N^{-1}(\rho_{\cal J}^{(L_1)}+ \rho_{\cal J}^{(L_2)}+\ldots,\rho_{\cal
     J}^{(L_N)})$, in the limit $N\to\infty$.  It can be proved
     \cite{NSBerlin,AW90} that $\rho_{\cal J}(\sigma)$ is indeed a Gibbs state.

     One can also construct overlaps without constructing Gibbs states at all,
     as is done numerically.  Such a construction (similar to that above) is
     described in \cite{NS96c}, and leads ultimately to the same conclusion.

     It is easy to show, given the torus-translation symmetry inherent in
     periodic b.c.'s, that the Gibbs state $\rho_{\cal J}(\sigma)$ is
     translation-{\it covariant\/}; that is, $\rho_{{\cal
     J}^a}(\sigma)=\rho_{\cal J}(\sigma^{-a})$, or in terms of correlations,
     $\langle\sigma_x\rangle_{{\cal J}^a}=\langle\sigma_{x-a}\rangle_{\cal J}$.
     Translation covariance of $\rho_{\cal J}$ immediately implies, via
     Eqs.~(\ref{eq:overlap})-(\ref{eq:PJ(q)}), translation invariance of
     $P_{\cal J}$.  But, given the translation-ergodicity of the underlying
     disorder distribution $\nu$, {\it it immediately follows that\/} $P_{\cal
     J}(q)$ {\it is self-averaging, and equals its distribution average $P(q)$
     for a.e.~\/}${\cal J}$.  The same result can be shown for other
     coupling-independent b.c.'s, where torus-translation symmetry is absent,
     using methods described in \cite{NS2D00}.  A simple argument, given in
     \cite{NS96c}, shows further that nontrivial ultrametricity, as in the
     Parisi solution of the SK model, cannot hold among three arbitrarily chosen
     states.

     So the most natural interpretation of the RSB picture cannot be applied to
     short-ranged spin glasses.  The question then becomes: are there
     alternative, less straightforward interpretations?

     \subsection{The Nonstandard SK Picture}
     \label{subsec:nonstandard}

     The standard SK picture follows a traditional approach in its focus on
     thermodynamic states.  We argued in Sec.~\ref{sec:metastates} that for
     inhomogeneous systems like spin glasses, such a focus is too restrictive
     if many pure states are present.  Instead, the metastate approach is far
     better suited as a guide for analyzing these kinds of systems.

     We have mentioned above (Sec.~\ref{subsubsec:observ}), and will describe in
     detail below, the issue of the overlap function $P(q)$ being a poor tool
     for determining numbers of pure states in short-ranged models.  For the
     moment, however, let's assume that this is not the case, and that numerical
     simulations on the EA model in three and higher dimensions detect a
     Parisi-like overlap structure (see, e.g., \cite{MPRRZ00}).  Does the
     interpretation given in the cited paragraph in Sec.~\ref{subsec:standard}
     necessarily follow?  The answer is {\it no\/}; it was shown in \cite{NS96b}
     that any evidence for RSB arising from numerical studies of $P(q)$ can
     correspond to more than one thermodynamic picture.  This leads to a
     reinterpretation not only of what broken replica symmetry might mean in
     short-ranged systems, but also what it {\it does\/} mean in the SK model.

     In numerical computations, overlaps are by necessity computed in finite
     volumes.  Because of CSD, it cannot be assumed that a simple $L\to\infty$
     extrapolation leads to a single thermodynamic mixed state whose
     decomposition includes all of the observable pure states of the system.
     But if one is indeed observing a nontrivial (i.e., decomposable into many
     pure states) mixed state $\Gamma$ in {\it one\/} volume, why should one
     expect that a similar observation in a {\it different\/} volume corresponds
     to the same $\Gamma$?  In fact, the presence of CSD indicates that this
     {\it cannot\/} be the case for all large volumes.

     As a mathematical aside, the standard SK picture effectively corresponds to
     breaking the replica symmetry {\it after\/} the thermodynamic limit has
     been taken.  Numerical studies, however, essentially break the replica
     symmetry {\it before\/} taking the thermodynamic limit.  Guerra
     \cite{Guerra} has noted that the order of these limits can be significant.
     That an interchange of such limits can lead to a new thermodynamic picture
     of the spin glass phase does not seem to have been appreciated prior
     to~\cite{NS96b}.

     Based on these considerations, a new, nonstandard interpretation of the
     mean-field RSB picture was introduced in \cite{NS96b} and described in
     detail in Sec.~VII of \cite{NS97}.  It is a maximal mean-field picture,
     preserving mean-field theory's main features, although in an unusual way.
     The most natural description of this nonstandard interpretation is in terms
     of the metastate.
       
       We now summarize, informally, the nonstandard SK picture.  Formal
       treatments can be found in \cite{NSBerlin,NS96b}; other detailed
       descriptions appear in \cite{NS02,NS98,NS97}.

       We again consider the PBC metastate, although, as always, almost any other
       coupling-independent metastate will suffice.  In order to resemble the
       structure of ordering in the SK model as closely as possible, the
       nonstandard SK picture assumes that in each $\Lambda_{L_i}$, the finite-volume
       Gibbs state $\rho_{{\cal J},{L_i}}$ is well approximated deep in the interior
       by a mixed thermodynamic state $\Gamma^{({L_i})}$, decomposable into many pure
       states $\rho_{\alpha_{L_i}}$:
       \begin{equation}
       \label{eq:gamma}
       \Gamma^{({L_i})}=\sum_{\alpha_{L_i}}W_{\Gamma^{(L_i)}}^{\alpha_{L_i}}\rho_{\alpha_{L_i}}\,
       .
       \end{equation}
       In this equation, explicit dependence on ${\cal J}$ is suppressed.
       $\Gamma^{({L_i})}$ is a thermodynamic mixed state decomposable into pure
       states $\rho_{\alpha_{L_i}}$; the index ${L_i}$ is meant only to indicate
       the volumes in which $\Gamma^{({L_i})}$ appears.

       Each mixed state $\Gamma^{(L_i)}$ is presumed to have a nontrivial overlap
       distribution
       \begin{equation}
       \label{eq:gamov}
       P_{\Gamma^{(L_i)}}=\sum_{\alpha_{L_i},\beta_{L_i}}W_{\Gamma^{(L_i)}}^{\alpha_{L_i}}
       W_{\Gamma^{(L_i)}}^{\beta_{L_i}}\delta(q-q_{\alpha_{L_i}\beta_{L_i}})
       \end{equation}
       of the form shown in Fig.~5.  Moreover, the distances among any three pure
       states {\it within a particular\/} $\Gamma$ are assumed to be ultrametric.

       As already noted, the presence of CSD requires that $\Gamma^{({L_i})}$ change
       in some ``chaotic'' fashion with ${L_i}$.  Hence, if one computes the overlap
       distribution in a {\it particular\/} $\Lambda_{L_i}$, one would see something
       like Fig.~5.  However, if one looks at two typical volumes of very
       different sizes, one would see something like Fig.~8.

       \begin{figure}
       \label{fig:nonstandard}
       \centerline{\epsfig{file=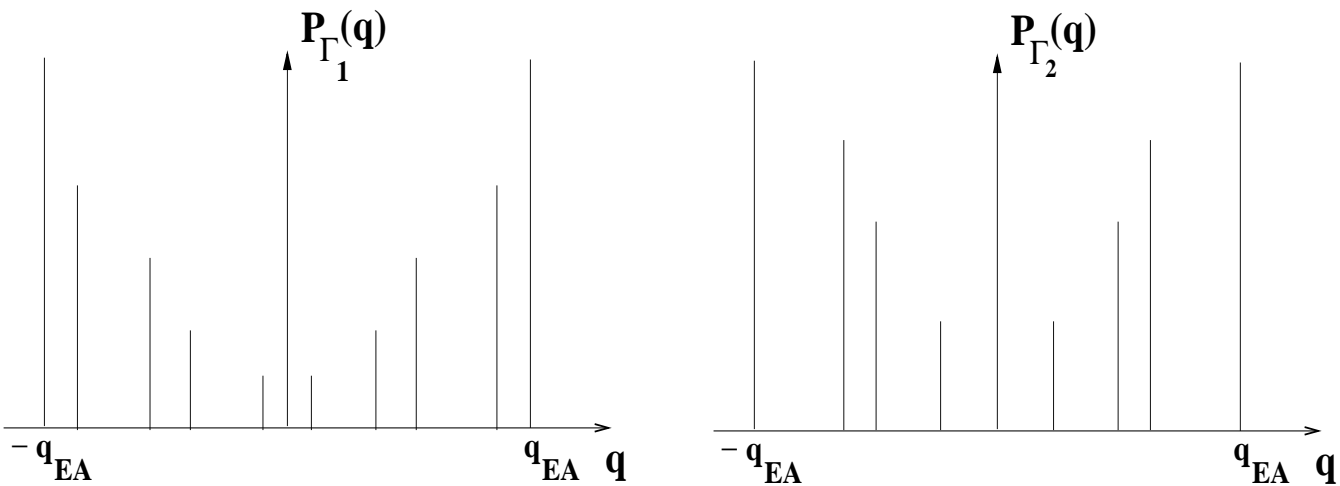,width=7.0in,height=2.5in}}
       \renewcommand{\baselinestretch}{1.0} 
       \small
       \caption{The overlap distribution, at fixed ${\cal J}$, in two different
       volumes $\Lambda_1$ and $\Lambda_2$ in the nonstandard SK picture.}
       \end{figure}
       \renewcommand{\baselinestretch}{1.25}
       \normalsize

       \subsection{What Non-Self-Averaging {\it Really\/} Means}
       \label{subsec:really}

       In this section we discuss why the nonstandard SK picture forces a
       redefinition of the meaning of non-self-averaging.  An important
       consequence of this redefinition is that most quantities of interest can
       now be defined for a {\it single\/} realization of the disorder: this
       includes the overlap distribution function.  So it is possible after all to
       focus on a particular sample rather than an ensemble of samples.

       The same argument, given in Sec.~\ref{subsec:standard}, that showed the
       translation-covariance of the thermodynamic state $\rho_{\cal J}(\sigma)$
       applies equally to the metastate $\kappa_{\cal J}$.  As a consequence, the
       resulting {\it ensemble\/} of overlap distributions $P_{\Gamma^{({L_i})}}$
       is independent of ${\cal J}$.  The dependence of the overlap distribution
       on $\Lambda_{L_i}$ (as $\Gamma^{({L_i})}$ varies within the metastate
       ensemble), no matter how large ${L_i}$ becomes, is the redefinition of
       non-self-averaging.  It replaces dependence of overlap distributions on
       ${\cal J}$ with dependence on ${L_i}$ for {\it fixed\/} ${\cal J}$.

       So, instead of averaging the overlap distribution over ${\cal J}$, the
       averaging {\it must now be done over the states\/} $\Gamma$ {\it within the
       metastate\/} $\kappa_{\cal J}$, all at fixed ${\cal J}$:
       \begin{equation}
       \label{eq:Pgamma}
       P(q)=[P_\Gamma(q)]_{\kappa_{\cal J}}=\int P_\Gamma(q)\kappa_{\cal
       J}(\Gamma)d\Gamma\, .
       \end{equation}
       The $P(q)$ thus obtained for a single ${\cal J}$ has the form shown in
       Fig.~6, and is the same for a.e.~${\cal J}$.

       \subsection{Differences Between the Standard and Nonstandard Pictures}
       \label{subsec:differences}

       The nonstandard SK picture differs from the usual one in several important
       respects.  One is the lack of dependence of overlap distributions on ${\cal
       J}$, and the replacement of the usual sort of non-self-averaging with that
       of dependence on states within the metastate.  Another important difference
       is that, in the nonstandard SK picture, a continuum (and therefore, an
       uncountable infinity) of pure states {\it and\/} their overlaps must be
       present; therefore, {\it ultrametricity would not hold in general among any
       three pure states chosen at fixed\/} ${\cal J}$, for the same reason
       ultrametricity breaks down in the standard SK picture \cite{NS96c}.
       Instead, the pure states at fixed ${\cal J}$ are split up into a continuum
       of families, where each family consists of those pure states occurring in
       the decomposition of a particular $\Gamma$, and only {\it within\/} each
       such family would ultrametricity hold.

       This has an important zero-temperature implication, because {\it there is
       no difference between the standard and nonstandard SK pictures at $T=0$.\/}
       The reason is that, for any finite $L$ and, say, periodic boundary
       conditions, there will be only a single ground state pair $\pm\sigma_0^L$
       in $\Lambda_L$.  (We assume as usual that the coupling distribution is
       continuous, such as Gaussian, to avoid accidental degeneracies.)  {\it It
       follows that overlaps of ground states cannot display nontrivial
       ultrametricity, or any other nontrivial structure.}

       We have presented the nonstandard SK picture as a replacement for the more
       standard mean-field picture; if realistic spin glasses display any
       mean-field features, something like it must occur.  However, this leaves
       open the question of what {\it does\/} happen in realistic spin glasses.
       In particular, can the nonstandard SK picture actually occur in
       short-ranged spin glasses? In the next section we show that the answer,
       again, is no.

       \subsection{Invariance of the Metastate}
       \label{subsec:invariance}

       In this section we present a theorem about the metastate whose proof is
       relatively simple but whose implications are powerful and far-reaching, not
       only for spin glasses but for disordered and inhomogeneous systems in
       general.  However, we restrict the discussion here to the EA model in any
       finite dimension and at any temperature, in zero field and with a symmetric
       coupling distribution.  We consider again flip-related boundary conditions,
       such as periodic and antiperiodic, or any two fixed b.c.'s.  

       \medskip

       {\bf Theorem 3 \cite{NS98}.}  Consider two metastates constructed along a
       deterministic sequence of $\Lambda_L$'s, using two different sequences of
       flip-related, coupling-independent b.c.'s.  Then with probability one,
       these two metastates are the same.

       \medskip

       The proof, given in \cite{NS98}, is relatively straightforward, and uses
       two ingredients.  The first is that, as proved in \cite{NSBerlin}, along
       some deterministic subsequence of volumes both the histogram construction
       of metastates and the Aizenman-Wehr construction \cite{AW90} result in the
       same metastate.  But because the latter method averages over couplings
       outside of each volume, it rigorously follows (exactly as in the proof of
       Theorem~2 in Sec.~\ref{subsubsec:manystates}) that two metastates constructed
       with flip-related b.c.'s must be the same.

       Despite the straightforward nature of the theorem and proof, it is a
       striking result, and its consequences for the nature of the spin glass
       state are immediate.  Not only are the periodic and antiperiodic metastates
       the same; if one were to choose, in a ${\cal J}$-independent manner, two
       {\it arbitrary\/} sequences of periodic and antiperiodic b.c.'s, the
       metastates (with probability one) would {\it still\/} be identical.  That
       is, the metastate, and corresponding overlap distributions constructed from
       it, at any fixed temperature and in any dimension are highly {\it
       insensitive\/} to boundary conditions.

       This invariance with respect to different sequences of periodic and
       antiperiodic b.c.'s means that the frequency of appearance of various
       thermodynamic states $\Gamma^{(L)}$ in finite volumes $\Lambda_L$ is (with
       probability one) {\it independent\/} of the choice of boundary conditions.
       Moreover, this same invariance property holds (with probability one) among
       any two sequences of {\it fixed\/} boundary conditions (and the fixed
       boundary condition of choice may even be allowed to vary arbitrarily along
       any single sequence of volumes)!  It follows that, with respect to changes
       of boundary conditions, the metastate is extraordinarily robust.

       This insensitivity would be unsurprising if there were only a single
       thermodynamic state, such as paramagnetic, or a single pair of flip-related
       states, as in droplet/scaling.  But it is difficult to see how our
       result can be reconciled with the presence of {\it many\/} thermodynamic
       states; indeed, at first glance it would appear to rule them out.

       However, while it does not rule out the possibility of many states,
       Theorem~3 does put severe constraints on the form of the metastate and its
       overlap distribution function.  In light of this strong invariance
       property, any metastate constructed via coupling-independent b.c's should
       be able to support only a very simple overlap structure, effectively ruling
       out the nonstandard SK picture.

       The nonstandard SK picture requires (cf.~Eq.~(\ref{eq:gamma})) that the
       $\Gamma$'s appearing in the metastate be of the form $\sum_\alpha
       W_\Gamma^\alpha\rho_\Gamma^\alpha$, with the weights $W_\Gamma^\alpha$ in
       each $\Gamma$ nonzero and unequal.  That implies that with periodic b.c.'s,
       say, the fraction of $L_i$'s for which the finite volume Gibbs state in
       $\Lambda_{L_i}$ puts (e.g.) at least 84\% of its weight in one pair of pure
       states is, say, 0.39.  But then it must also be the case that with antiperiodic
       b.c.'s the fraction of volumes for which the finite volume Gibbs state puts
       at least 84\% of its weight in some unspecified pair is still exactly 0.39!
       Moreover, the same argument must apply to any ``cut'' one might care to
       make; i.e., one constructs the periodic b.c.~metastate and finds that $x\%$
       of all finite volumes have put $y\%$ of their weight in $z$ states, with
       $z$ depending on the (arbitrary) choice of $x$ and $y$.  Then this must be
       true also for all volumes with antiperiodic b.c's; and similarly (but
       possibly separately) among all pairs of fixed b.c.~states.

       The only sensible way in which this could happen would be for the selection
       of states to be insensitive to the choice of boundary conditions, i.e., a
       particular sequence of b.c.'s should not prefer any states over any others,
       so that $\rho_{\cal J}$, the average over the metastate, would be some sort
       of uniform mixture of the pure states.  However, this {\it cannot\/} happen
       when the $\Gamma$'s are nontrivial mixed states, as in
       Eq.~(\ref{eq:gamma}).  

       The essential reason for this is that the weights in Eq.~(\ref{eq:gamma})
       must change with ${\cal J}$, as seen heuristically by the following
       argument.  Choose a particular coupling $J_{xy}$ and consider the
       transformation $J_{xy}\to J'_{xy}=J_{xy}+\Delta J$.  Then the weight
       $W^\alpha$ of the pure state $\alpha$ within $\Gamma$ will transform as
       \begin{equation}
       \label{eq:walpha}
       W^\alpha\to W'^\alpha=r_\alpha W^\alpha/\sum_\gamma r_\gamma W^\gamma
       \end{equation}
       where 
       \begin{equation}
       \label{eq:ralpha}
       r_\alpha=\left\langle\exp(\beta\Delta J\sigma_x\sigma_y)\right\rangle_\alpha=
       \cosh(\beta\Delta J) +
       \langle\sigma_x\sigma_y\rangle_\alpha\sinh(\beta\Delta J)\ ,
       \end{equation}
       for each pure state $\alpha$.  

       Now if each $\Gamma$ has a nontrivial decomposition over many pure state
       pairs, as in Eq.~(\ref{eq:gamma}), then the different pure state pairs
       differ in at least some even correlation functions; i.e., they have
       relative domain walls.  Eqs.~(\ref{eq:walpha}) and (\ref{eq:ralpha}) would
       then lead to a change of the relative weights of the different pure state
       pairs.  As a consequence, even a finite change in a {\it single\/} coupling
       $J_{xy}$ would change a uniform mixture (e.g., in the metastate average) to
       a non-uniform one.  But, of course, the mixture must be the same for
       a.e.~${\cal J}$.

       To summarize, the invariance of the metastate with respect to boundary
       conditions appears to be inconsistent with the transformation properties of
       $\Gamma$'s of the form Eq.~(\ref{eq:gamma}) with respect to finite changes
       in ${\cal J}$.  This leads to a contradiction, ruling out not only
       nonstandard SK but {\it any picture\/} in which the $\Gamma$'s are a
       nontrivial mixture of pure states.  When combined with our previous
       elimination of more standard versions of the mean field picture, it removes
       the possibility of {\it any\/} version of mean field ordering in
       short-ranged spin glasses.

       The invariance property of the metastate requires that both the pure state
       structure and the overlap structure of realistic spin glasses should be
       relatively simple.  So what are the remaining, plausible possibilities for the
       structure of the spin glass phase?  In the next section, we will introduce
       a new picture, guided once again by metastate concepts.

       \section{Structure of the Low-$T$ Spin Glass Phase}
       \label{sec:structure}

       \subsection{Remaining Possibilities}
       \label{subsec:possibilities}

       Assuming that spin-flip symmetry is broken below some $T_c(d)>0$, with $d$
       greater than some lower critical dimension $d_c$, what then are the
       possible structures that the spin glass phase can assume?

       A two-state picture, in which the only observable pure states are global
       flips of each other, remains completely consistent with all rigorous
       results described in the preceding section.  As noted in
       Sec.~\ref{sec:open}, the droplet/scaling scenario is such a two-state
       picture.  At the same time, it should be emphasized that droplet/scaling
       makes additional assumptions that our results do not address.  If it were
       to be proved that there exists only a single pair of pure states in the
       spin glass phase in any finite dimension above $d_c$, this would lend
       strong support to droplet/scaling, but such a proof alone would not be
       sufficient to demonstrate its correctness.  For fuller discussions of
       droplet/scaling, see \cite{Mac84,BM85,BM87,FH86,HF87a,FH87b,FH88}.

       Are there any many-state pictures that {\it are\/} consistent with the
       strong constraints imposed by the invariance of the metastate?  There is
       such a picture, introduced by the authors in \cite{NS92}, and discussed in
       detail in \cite{NSBerlin,NS02,NS98,NS96b,NS97}.  It is called the {\it
       chaotic pairs\/} picture, for reasons that will become apparent below.

       In a simple two-state picture like droplet/scaling, the overlap
       distribution function $P_{\cal J}^L(q)$ in a volume $\Lambda_L$ will simply
       approximate a sum of two $\delta$-functions, as in Fig.~4, but with $\pm
       M^2(T)$ replaced by $\pm q_{EA}$.  In the chaotic pairs picture, each
       finite-volume Gibbs state $\rho^L_{\cal J}$ will still be approximately a
       mixture of a {\it single\/} pair of spin-flip-related pure states, {\it but
       now the pure state pair will vary chaotically with $L$.\/} Then for each
       $\Lambda_L$, $P_{\cal J}^L(q)$ will again approximate a sum of two
       $\delta$-functions at $\pm q_{EA}$.  This picture is fully consistent with
       metastate invariance.

       So chaotic pairs resembles the droplet/scaling picture in finite volumes,
       but has a very different thermodynamic structure.  It is a many-state
       picture, but unlike the mean-field picture, only a {\it single\/} pair of
       spin-reversed pure states $\rho_{\cal J}^{\alpha_L}$, $\rho_{\cal
       J}^{\overline{\alpha_L}}$, appears in a large volume $\Lambda_L$ with
       symmetric boundary conditions, such as periodic.  In other words, for large
       $L$, one finds that
       \begin{equation}
       \label{eq:possfive}
       \rho_{\cal J}^{(L)}\approx {1\over 2}\rho_{\cal J}^{\alpha_L}+{1\over
       2}\rho_{\cal J}^{\overline{\alpha_L}}\, .
       \end{equation}
       Here, the pure state pair (of the infinitely many present) appearing in
       finite volume $\Lambda_L$ depends chaotically on $L$.  Unlike the
       droplet/scaling picture, this new possibility exhibits CSD with periodic
       b.c.'s.  So, like nonstandard SK, the periodic b.c.~metastate is
       dispersed over (infinitely) many $\Gamma$'s, but unlike nonstandard SK each
       $\Gamma$ is a {\it trivial\/} mixture of the form $\Gamma =
       \Gamma^\alpha={1\over 2}\rho_{\cal J}^\alpha+{1\over 2}\rho_{\cal
       J}^{\overline{\alpha}}$.  The overlap distribution for each $\Gamma$ is the
       same: $P_\Gamma = {1\over 2}\delta(q-q_{EA})+{1\over 2}\delta(q+q_{EA})$.
       It is interesting to note that the highly disordered spin glass model
       \cite{NS94,NS96a,BCM94}, mentioned in Sec.~\ref{subsec:frustration},
       appears to display just this behavior in its ground state structure in
       sufficiently high dimension.

       Why doesn't an argument similar to that used to rule out nonstandard SK
       also rule out chaotic pairs?  Because in the chaotic pairs picture, as in
       droplet/scaling, there are in each $\Gamma$ only two pure states (depending
       on $\Gamma$ in chaotic pairs), each with weight $1/2$.  All even
       correlations are the same in any pair of flip-related pure states, so, by
       Eqs.~(\ref{eq:walpha}) and (\ref{eq:ralpha}), any change in couplings
       leaves the weights unchanged.

       There is an interesting additional piece of information that metastate
       invariance supplies for many-state pictures like chaotic pairs: the number
       of pure state pairs (if infinite) must be an {\it uncountable\/} infinity.
       If there were a {\it countable\/} infinity, one couldn't have a uniform
       distribution consisting of all equal, positive weights within the
       metastate.

       The term ``chaotic pairs'' was chosen in reference to spin-symmetric
       b.c.'s, such as periodic.  If one considers a fixed b.c.~metastate, then it
       would be more appropriate to refer to this picture as ``chaotic pure
       states'', because the Gibbs state in a typical large volume $\Lambda_L$
       with fixed b.c.'s will be (approximately) a single pure state that varies
       chaotically with $L$.  But the thermodynamic picture is the same, and just
       manifests itself slightly differently in different metastates.  In
       particular, the {\it average\/} over the metastate $\rho_{\cal J}$
       (cf.~Eq.~(\ref{eq:rhoav})) should be the same for periodic and fixed b.c.'s.

       We conclude with a brief note about zero temperature, where we observed in
       Sec.~\ref{subsec:differences} that in each $\Lambda_L$ there is only a
       single ground state pair.  If droplet/scaling holds, then this pair is the
       same for all large $L$; if infinitely many ground state pairs exist, then
       the pair changes chaotically with $L$.  This will be true at $T=0$ for {\it
       any\/} many-state picture, whether chaotic pairs, mean-field RSB, or some
       other such picture.  The metastates, hence overlap functions, of these
       many-state pictures differ only at positive temperature: the mean-field RSB
       picture at $T>0$ consists of the $\Gamma$ in each volume exhibiting a
       nontrivial mixture of pure state pairs as in Eq.~(\ref{eq:gamma}), while in
       chaotic pairs the $\Gamma$ appearing in any $\Lambda_L$ consists of a
       single pure state pair, as in Eq.~(\ref{eq:possfive}).  So the periodic
       b.c.~metastate in the chaotic pairs picture looks similar at zero and
       nonzero temperatures, like droplet/scaling but unlike nonstandard SK.

       \subsection{The Problem with $P(q)$}
       \label{subsec:problem}

       In Sec.~\ref{subsec:nonstandard}, we discussed the usual numerical
       procedure for constructing overlaps.  In both chaotic pairs and
       droplet/scaling, the overlap distribution function in almost every large
       volume (computed in a window far from the boundaries, to avoid nonuniversal
       boundary effects) is simply a single pair of $\delta$-functions at $\pm
       q_{EA}$, as in Fig.~4.  {\it So using $P(q)$ in the usual way cannot
       differentiate between a many-state picture like chaotic pairs and a
       two-state picture like droplet/scaling\/}.

       One could try an alternative approach, such as looking at the average state
       over many volumes, as an approximation to studying the average over the
       metastate $\rho_{\cal J}(\sigma)$.  This average looks very different in
       the two pictures: in droplet/scaling, it is still a trivial mixture of two
       states, of the form Eq.~(\ref{eq:possfive}), while in chaotic pairs, it is
       presumably a uniform mixture over uncountably many states.  This approach,
       in effect, takes the thermodynamic limit {\it before\/} breaking the replica
       symmetry, as discussed in Sec.~\ref{subsec:nonstandard}.

       Now there {\it will\/} be a difference between overlap functions in
       droplet/scaling and chaotic pairs.  In droplet/scaling, $P_{\cal J}(q)$ is
       again a pair of $\delta$-functions at $\pm q_{EA}$.  In chaotic pairs,
       $P_{\cal J}(q)$ would now most likely equal $\delta(q)$: it was proven in
       \cite{NS99a}, and will be discussed in Sec.~\ref{subsec:metastable}, that
       $P_{\cal J}(q)=\delta(q)$ for the spin overlaps of $M$-spin-flip-stable
       metastable states for any finite $M$.  If there are infinitely many ground
       state pairs, we expect the same to be true for ground states, i.e., for
       $M=\infty$.  But, although there is now a difference between overlap
       functions in droplet/scaling and chaotic pairs, there is now also {\it
       no\/} difference between overlap functions in chaotic pairs and in the
       simple paramagnet!

       So, although the form of the overlap function can depend on how its
       computation is done, {\it the overlap structure in spin glasses must be
       simple\/}, regardless of whether there are infinitely many pure states or
       only a single pair.  Moreover, the overlap function cannot distinguish
       between many states, and one or two states, in an unambiguous manner.

       Similar problems with overlap functions, in particular their sensitivity to
       boundary conditions for short-ranged systems, are discussed in
       \cite{HF87a}. There it is noted that in the three-dimensional random field
       Ising model at low temperature and weak field magnitude, $P_L(q)$ computed
       in a given $\Lambda_L$ will miss one of the relevant pure states.
       Conversely, in the two-dimensional Ising ferromagnet on a square lattice,
       at low temperature, and with antiperiodic b.c.'s in both directions,
       $P_L(q)$ gives the appearance of many states where there are only two.

       A final, interesting example was suggested to us by A.~van~Enter and
       appears in the Appendix of \cite{NS97}.  Consider $P_L(q)$ for an Ising
       antiferromagnet in two dimensions with periodic boundary conditions.  For
       {\it odd\/}-sized squares $P_L(q)$ is the same as that of the ferromagnet
       with {\it periodic\/} boundary conditions, and for {\it even\/}-sized
       squares it is equivalent to that of the ferromagnet with {\it
       antiperiodic\/} boundary conditions.  The overlap distribution computed in
       the full volume therefore oscillates between two different answers, an
       example of CSD for {\it overlap\/} distributions.  This illustrates the
       importance of computing quantities in windows that are much smaller than
       the system size.  In this case, restricting $P(q)$ to such a window gives
       rise to a well-defined answer: a pair of $\delta$-functions.

       \section{Interfaces}
       \label{sec:interfaces}

       Our discussion up to now has focused largely on the numbers and
       organization of pure states in the spin glass phase, if it exists.  Even
       though these are fundamental constructs from the point of view of
       thermodynamics, and their multiplicity and organization directly affect
       observable equilibrium \cite{BY86} and dynamical \cite{NS99b} spin glass
       properties, they may still seem somewhat abstract.  In this section we will
       draw on recent results that provide a concrete link between the structure
       of interfaces on the one hand, and numbers of pure states, the
       relationships among them, and their low-lying excited states on the other.
       At the same time, we will extend and clarify our earlier distinction
       (Sec.~\ref{subsubsec:observ}) between observable and invisible states.

       \subsection{Space-Filling Interfaces and Observable States}
       \label{subsec:spacefilling}

       In Sec.~\ref{subsubsec:free}, we introduced the notion of ``space-filling''
       interfaces; these play a central role in what follows.  The {\it
       interface\/} between two spin configurations $\sigma^L$ and ${\sigma'}^L$
       in $\Lambda_L$ is simply the set of couplings satisfied in one and not the
       other spin configuration; a domain wall is a connected component of an
       interface.  So the interface between two configurations is the union of all
       domain walls, and may consist of one or many.  We usually envision domain
       walls as lines or surfaces in the {\it dual\/} lattice, cutting those bonds
       satisfied in one but not the other spin configuration.  Domain walls
       therefore separate regions (in the real lattice) in which spins in
       $\sigma^L$ and ${\sigma'}^L$ agree from regions where they disagree.

       We will confine our remarks here to zero temperature, although it is
       possible to extend the discussion to nonzero temperatures.  We are
       interested only in interfaces whose linear spatial extent $l$ is $O(L)$ 
       in $\Lambda_L$.  At zero temperature, these are the {\it only\/}
       interfaces between ground state configurations, if the coupling
       distribution is continuous; that this is so follows from the same argument
       proving that domain walls separating infinite spin configurations cannot
       end in any finite region and cannot have closed loops \cite{NS00}.
       However, the restriction against closed loops need not apply to interfaces
       separating ground and excited states.

       If the number of couplings in an interface of linear extent $l$ scales as
       $l^{d_s}$, we call $d_s$ the dimension of the interface. A {\it
       space-filling\/} interface is one with $d_s=d$.  One of the interesting
       features of spin glasses is the possibility of ground or pure states
       separated by space-filling interfaces, unlike in ferromagnets where $d_s<d$
       always.  For example, the interface ground states generated from Dobrushin
       boundary conditions (all boundary spins on one side of a plane bisecting
       $\Lambda_L$ fixed at $+1$ and the remaining boundary spins $-1$) have a
       single domain wall relative to the uniform ground states, with $d_s=d-1$.

       The first result of this section is a theorem first proved in \cite{NS00}
       indicating that interfaces separating {\it observable\/} ground states are
       space-filling.

       \medskip  

       {\bf Theorem 4 \cite{NS00}.} Suppose that in the EA model in some finite
       dimension $d$, there exists more than a single pair of thermodynamic ground
       states in a coupling-independent metastate.  Then the interface between 
       two ground states chosen from different pairs must have $d_s=d$.

       \medskip

       {\bf Proof.}  We sketch the proof here for periodic boundary condition
       metastates; the extension to other coupling-independent metastates can be
       obtained using the procedure presented in \cite{NS01a}.  Consider the
       periodic b.c.~metastate $\kappa_{\cal J}$.  By taking two ground state
       pairs chosen randomly from $\kappa_{\cal J}$, one obtains a configuration
       of interfaces.  This procedure yields a measure ${\cal D}_{\cal J}$ on
       domain walls.  By integrating out the couplings, one is left with a
       translation-invariant measure ${\cal D}$ on the domain wall configurations
       themselves.  By the translation-invariance of ${\cal D}$, any
       ``geometrically defined event'', e.g., that a bond belongs to a domain
       wall, either occurs nowhere or else occurs with strictly positive
       density. This immediately yields the result. $\diamond$

       \medskip

       {\it Remark.\/} Theorem~4 extends to pure states at nonzero
       temperatures.

       \subsection{Invisible States}
       \label{subsec:invisible}

       Our emphasis in this review has been on observable pure states, which we
       have identified as the `physical' states; that is, we expect only those
       states to influence outcomes of laboratory measurements or numerical
       simulations.  Still, it may be interesting to consider also the
       characteristics of `invisible' states, should they be present.

       Recall that an invisible pure state is one that does not lie in the support
       of any coupling-independent metastate.  Invisible
       pure states can only be generated by sequences of coupling-{\it
       dependent\/} boundary conditions.  There is presently no known method for
       constructing such states, and it is not clear whether such constructions,
       if found, would be measurable (Sec.~\ref{sec:csd}). Consequently, it is not
       presently known whether such states actually exist in spin glasses, or if
       they do, whether they would be of any interest other than mathematical.
       Nevertheless, if they should be found to exist, one can still make some
       predictions about their properties.

       From the discussion in Sec.~\ref{subsubsec:free}, the interfaces
       between invisible states would necessarily have energies scaling as
       $l^\theta$, with $\theta>(d-1)/2$; any interface with lower energy
       would almost certainly show up in a coupling-independent metastate.
       So a ``high-energy'' interface, satisfying the above condition,
       would separate invisible pure states from observable ones (or each
       other), regardless of the interface dimensionality \cite{notevE}.

       It also follows, by Theorem~4, that any interface with dimension $d_s<d$
       separating pure states would necessarily appear only from
       coupling-dependent boundary conditions.  Such an interface would
       necessarily have high energy; if its energy exponent $\theta\le (d-1)/2$,
       it would not correspond to an interface between invisible and observable
       pure states, but rather to one between excited and ground states.  This is
       discussed further in Sec.~\ref{subsec:excitations}.

       \subsection{Relation Between Interfaces and Pure States}
       \label{subsec:relation}

       We conclude from the previous discussion that interfaces separating
       observable ground (or pure) states are both space-filling and have energies
       scaling as $l^\theta$, where $0\le\theta\le (d-1)/2$.  In \cite{NS02} we
       proved the converse, namely that the existence of such interfaces is a
       sufficient condition for the existence of more than one thermodynamic pure
       state pair.  Specifically, the presence of space-filling interfaces is
       already sufficient for the existence of multiple pure state pairs; the
       energy condition is necessary for the pure states to be observable.

       As noted in Sec.~\ref{subsubsec:free}, RSB theory predicts that interfaces
       between ground states in $\Lambda_L$ (say, under a switch from periodic to
       antiperiodic boundary conditions) are both space-filling and have energies
       of $O(1)$ independent of lengthscale \cite{MP00a,MP00b,PY00,KM00}; i.e.,
       they have exponent $\theta=0$.  By the above theorem, the space-filling
       property requires the existence of multiple thermodynamic pure states, and
       the $O(1)$ energy property implies that typical large $\Lambda_L$'s with,
       say, periodic boundary conditions, would exhibit thermodynamic states that
       are nontrivial mixtures of different pure state pairs (see also
       \cite{NS98}).  But in Sec.~\ref{subsec:invariance} it was shown that such
       states cannot appear in the EA model.  It follows that space-filling
       interfaces with $O(1)$ energy are ruled out.

       \subsection{Low-Lying Excited States}
       \label{subsec:excitations}

       The discussion of the preceding two sections was restricted to pure or
       ground states.  One can construct other types of metastates, such as
       ``excitation metastates'' \cite{NS01a}, whose support includes both ground
       and low-lying excited states.  One example of an excitation metastate is
       the ``uniform perturbation metastate'' \cite{NS01b}, which we now describe.

       Once again, consider a deterministic sequence $\Lambda_L$ of cubes with
       periodic boundary conditions; at zero temperature, each such cube has a
       single pair $\pm\sigma_0^L$ of ground states.  For each $L$, consider a
       second spin configuration $\sigma'^L$, generated by some prespecified
       procedure (examples will be given below).  Then do for the {\it pair\/}
       $(\sigma_0^L,\sigma'^L)$ what was done for $\sigma_0^L$ in the original
       metastate; i.e., measure the relative frequency of occurrence of each pair
       (inside a fixed window, as always).  The resulting uniform perturbation
       metastate gives for both infinite-volume ground and excited states their
       relative frequency of appearance inside large volumes.

       The uniform perturbation metastate is useful when considering recent
       numerical studies \cite{PY00,KM00,HKM00,KPY01} on EA Ising spin glasses in
       three and four dimensions that may have uncovered new types of states.  In
       each volume, their interface with the ground state has $d_s<d$ with an
       energy of $O(1)$, independently of lengthscale.  (It should be noted that
       questions have been raised over the correct interpretation of these
       numerical results \cite{MP00a,Middleton00}, so we consider these states
       only as an interesting possibility.)  Because the two new constructions
       that lead to these states are done in a translation-invariant manner, a
       simple extension of Theorem~4 to uniform perturbation metastates
       \cite{NS01b} rules out the possibility that these new states can be ground
       or pure states; if they exist at all, they must be excited states.

       Although the two procedures are different, they seem to lead to similar
       outcomes.  The Krzakala-Martin \cite{KM00} procedure forces a random pair
       of spins $(\sigma_z,\sigma_{z'})$ to assume a relative orientation opposite
       to that in the ground state pair $\pm\sigma_0^L$; the rest of the spins are
       then allowed to relax to the available lowest-energy configuration.  This
       ensures that at least some bonds in the excited spin configuration,
       $\sigma'^L$, must be changed (i.e., satisfied $\leftrightarrow$
       unsatisfied) from $\sigma_0^L$.  It also ensures that the energy of
       $\sigma'^L$ is no more than $O(1)$ above that of $\sigma_0^L$, regardless
       of $L$.  In the Palassini-Young \cite{PY00} procedure, a novel
       coupling-dependent bulk perturbation ${\cal H}_{PY}$ is added to the
       Hamiltonian (\ref{eq:EA}) in $\Lambda_L$, where
       \begin{equation}
       \label{PY}
       {\cal
       H}_{PY}=(\epsilon/N_b)\sum_{<x,y>\in\Lambda_L}(\sigma_0^L)_x(\sigma_0^L)_y
       \sigma_x\sigma_y \ ,
       \end{equation}
       $\epsilon$ is independent of $L$, and $N_b$ is the number of bonds
       $<x,y>$ in $\Lambda_L$.  So here too the energy of
       $\sigma'^L$ is no more than $O(1)$ above that of $\sigma_0^L$.

       Krzakala-Martin and Palassini-Young excitations have $d_s<d$ and
       $\theta=0$.  In the two-state droplet picture of Fisher and Huse, however,
       excitations have $d_s<d$ and $0<\theta\le (d-1)/2$.  One focus of current
       research is to determine which, if either, of these pictures holds in
       realistic spin glasses.

       To summarize, spin configurations with space-filling interfaces
       correspond to new ground or pure states, observable if their energy
       exponent $\theta\le (d-1)/2$ and invisible if $\theta>(d-1)/2$.  If
       the ``space-filling'' exponent $d_s<d$, then the corresponding
       configurations {\it could\/} correspond to new ground or pure states
       only if $\theta>(d-1)/2$.  If $d_s<d$ and $\theta\le (d-1)/2$, such
       states can only be excitations \cite{NS01b}, 
       and do not signify the presence of additional pure states (for
       further discussion, see Sec.~\ref{subsubsec:purestates} below).

\section{Summary and Discussion}
\label{sec:discussion}

In this section we provide a brief summary of our conclusions and
discuss how they tie in to various other approaches.

\subsection{Summary}
\label{subsec:summary}

The central theme of this review is that a new set of concepts and methods
are needed in the treatment of statistical models with both disorder and
frustration.  While we have mostly focused on the strong disconnect between
infinite- and short-ranged spin glass models, we emphasize that this
conclusion essentially results from an application --- although a very
important one --- of our ideas and techniques.  These results should be
viewed as part of a more extensive and general framework for approaching
the study of disordered systems.  The unifying concept is that of the {\it
metastate\/} (Sec.~\ref{sec:metastates}), which broadens the focus of study
from the conventional one of thermodynamic states, to {\it distributions\/}
(or probability measures) on thermodynamic states --- i.e., {\it
ensembles\/} of thermodynamic states.  Within this framework, universal
themes that clarify the study of disordered systems can be considered: the
detectability of many states, chaotic size dependence, the relations
between interface types and pure states, `windows' in numerical
experiments, and others.

As noted, a prominent application of these methods leads to the conclusion
that the spin glass differs from most other statistical mechanical systems
in that its infinite-ranged version displays unique low-temperature
properties, not shared by corresponding short-ranged models in any
dimension and at any temperature (for other differences between homogeneous
and disordered systems in general, see Sec.~\ref{subsec:statmech}).  A
likely underlying cause of this difference is the combination of the
following features in the infinite-ranged spin glass: the presence of both
ferromagnetic and antiferromagnetic couplings, the statistical independence
of all of the couplings, and the scaling of their magnitudes to zero as
$N\to\infty$ (cf.~\cite{NS03} and Sec.~\ref{subsubsec:states}).  The
simultaneous presence of all three of these properties seems essential.  In
other systems (for example, a number of combinatorial optimization
problems) with analogous features, one should expect similar behavior.

In the following subsections, we expand upon these remarks.  We begin by
clarifying some essential issues.  Although the contents of the following
two subsections can be placed naturally into Sections~\ref{subsec:pure} and
\ref{subsec:standard}, respectively, the issues they treat have generated 
sufficient confusion that it seems worthwhile to address them separately.

\subsubsection{Are the Pure States We Discuss the `Usual' Ones?}
\label{subsubsec:purestates}

The spin glass literature is replete with well-defined terms like `pure
state' used in imprecise ways, and often interchangeably with other terms
such as `valley'.  A careful discussion has been given in
\cite{vEvH84}, to which we refer the reader.  We note that the definition
of pure states given in Sec.~\ref{subsec:pure} does in fact correspond, at
least on a heuristic level, to the intuitive notion of `valley' as a
collection of configurations separated from all other configurations by
barriers that diverge in the thermodynamic limit.

This last statement can be made more precise, as in \cite{vEvH84}, by
considering a specific dynamics; in our case, the dynamics could be the
lattice animal $M$-spin-flip dynamics discussed in
Sec.~\ref{subsubsec:method}, for any finite $M$.  Then a pure state may be
thought of as a collection of configurations that can be reached from each
other via the dynamics in finite time.  At a given dimension and
temperature, there may be only a single pure state or multiple, disjoint
pure states.

For an $N$-spin system, with $N$ finite, it is often useful to think of
spin configurations as belonging to different `pure states' if the
dynamical pathway (here it is necessary that $M\ll N$) connecting them
requires a time that scales as the exponent of $N$ to some power (see, for
example, \cite{MY82}).  This notion has been highly successful in
constructing dynamics-based solutions to the SK model
\cite{SZ81,Somp81,SZ82}.

\subsubsection{Is the $P_{\cal J}(q)$ Used to Rule Out the Standard SK Picture the `Correct' One?}
\label{subsubsec:correct}

There are two natural objections that are sometimes raised to the
conclusion of \cite{NS96c}, in which the standard SK picture is ruled out
(see, for example, \cite{Parisi96}).  The specific overlap distributions
constructed in \cite{NS96c} and described above involve certain types of
averaging, for example, over lengthscale for fixed ${\cal J}$.  Are these
in fact different from the $P_{\cal J}(q)$ described by RSB theory, or
studied in numerical experiments?  Moreover, might it not be that the
averaging process used in \cite{NS96c} is itself the cause of the
self-averaged nature of the resultant $P_{\cal J}(q)$?

The answer to these objections appears in \cite{NSreply}.  The first point
is that, because of chaotic size dependence, the presence of many states
probably forbids the construction of any (infinite-volume) $P_{\cal J}(q)$
with the following two properties: (i) the construction can be defined for
a.e.~${\cal J}$ (necessary if {\it any\/} sort of averaging over coupling
realizations is to make sense), and (ii) the construction entails no
averaging of any kind.  So, for example, the construction proposed in
\cite{Parisi96}, where $P_{\cal J}(q)$ is first defined for finite $L$,
followed by taking a straightforward $L\to\infty$ limit, cannot work ---
there is no limiting $P_{\cal J}(q)$ for this construction (if there are
many states).

However, even if there were a construction with the above two properties,
the conclusion would be unchanged, because {\it how $P_{\cal J}(q)$ is
constructed is irrelevant\/} --- there cannot exist any $P_{\cal J}(q)$
which is both a physical infinite volume object {\it and\/} which is
non-self-averaging.  In fact, the only way out would be to construct a
clearly unphysical $P_{\cal J}(q)$ that depends on the choice of origin of
the coordinate system.  This is an immediate (and rigorous) consequence of
the spatial ergodicity of the underlying disorder distribution, as
explained in \cite{NS96c}.

\subsection{Comparison to Other Work}
\label{subsec:other}

Our conclusions regarding the disconnect between the SK and EA models are
supported by rigorous results.  Nevertheless, there have been numerous
studies that claim to support the basic features of the replica symmetry
breaking, mean-field pure state structure in EA models in finite
dimensions.  In this section we discuss some of these studies and examine
how they can be reconciled with the theorems reviewed above.  Given the
large number of such studies, we confine ourselves to a small but
representative sample.  For conciseness' sake, we will not review studies
that claim to support pictures that are consistent with our results
\cite{MBK98}, such as droplet/scaling.  (We emphasize once again, however,
that our results make no claim about the correctness of droplet/scaling,
and are equally compatible with some other competing pictures.)

\subsubsection{Numerical Studies}
\label{subsubsec:numerical}

There have been numerous simulations of the EA model, mostly in three and
four dimensions, that claim to find results consistent with the RSB
picture.  We will restrict ourselves here to those studies that examine
equilibrium properties.  The number of such studies is still quite large,
and even with this restriction, we will not attempt to cover them all here;
an extensive review is given in \cite{MPRRZ00}.  A reasonably
representative sample is given by
\cite{MPR94,MPR97,CPPS90,RBY90,MPRR96,MPRR98}.

Most of these studies directly measure the spin overlap $P(q)$ (and in some
cases, also edge overlaps) for relatively small samples, and generally at
temperatures that are not too low.  Some problems associated with using
$P(q)$ have already been discussed in Sec.~\ref{subsec:problem}.  At least
as important, most of these studies measure the overlap in the {\it full\/}
volume under consideration, rather than in a much smaller `window'.  But as
noted in Sec.~\ref{subsubsec:windows}, boundary effects --- especially on
these relatively small samples --- will almost certainly dominate, so that
no reliable conclusions can be drawn about whether one is really ``seeing''
the pure state structure.  An illuminating example of how false conclusions
can be drawn in such a case is given in Sec.~VI of \cite{NS98}.

A further problem concerns the temperature at which many of these studies
is carried out.  As discussed in Sec.~5 of \cite{NS02}, thermal effects can
contribute heavily to nontrivial structure in either spin or edge overlaps,
and can therefore also lead to misleading conclusions.  Some suggestions
for overcoming these problems were presented in \cite{NS02}.

Given all of these issues, it is unclear exactly what is actually being
measured in these simulations, and extreme caution needs to be exercised
before any conclusions addressing the relevance of RSB to short-ranged
systems can be drawn from these studies.

\subsubsection{Analytical Studies}
\label{subsucsec:analytical}

There have been comparatively few analytical studies that try to extend
directly the RSB picture to short-ranged spin glass models.  An early
attempt examined the averaged free energy in a $1/d$ expansion about
$d=\infty$ \cite{GMY90}, and found thermodynamics consistent with those of
the SK model.  Near the critical temperature, an enhancement of RSB effects
was found.

A related but more detailed study was done by DeDominicis, Kondor, and
Temesv\'ari \cite{DKT97} who used a field-theoretical approach to extend
the RSB theory to short-ranged models via a one-loop expansion.  Here, the
expansion was done (at fixed dimension $d$) in $1/z$ about $z=\infty$,
where $z$ is the coordination number.  Above six dimensions, the leading
mean-field term does not appear to be significantly modified by the loop
corrections.

While an interpretation of the $1/d$ or $1/z$ expansion results that
supports the standard SK picture is ruled out by our own results, these
calculations may nevertheless indicate interesting behavior.  It should
first be noted, though, that the calculations, though extensive, involve a
number of uncontrolled approximations that need to be understood before any
conclusions can be reached.  One issue to resolve is the validity of an
expansion done about a singular solution ($d=\infty$ for simple
nearest-neighbor hypercubic lattices in \cite{GMY90}, and $z=\infty$ at
fixed dimension $d$ in \cite{DKT97}). It is also unclear what would happen
if one went to higher orders in perturbation theory, and even whether these
series converge.

Another interesting possibility was raised by M.A.~Moore \cite{M03}, who
noted that the standard RSB approach expands about only one of the saddles
of the mean-field solution.  However, experience with other systems (the
disordered ferromagnet treated using replica methods \cite{BMMRY87}, and
the randomly diluted ferromagnet treated using non-replica methods
\cite{McK94}) implies that one may need to consider all saddle points,
including nominally sub-dominant ones, to get the correct result.

\subsubsection{Renormalization Group Studies and Types of Chaoticity}
\label{subsubsec:chaos}

Our emphasis on the centrality of the connection between chaotic size
dependence and the presence of many states leads naturally to questions
regarding whether our result is connected in some way to the presence of
other types of chaos that have been encountered in earlier spin glass
studies.  One well-known example arose from studies \cite{MB1,MB2} of
Migdal-Kadanoff renormalization-group transformations on frustrated Ising
systems on hierarchical lattices.  In these studies, chaotic
renormalization-group trajectories were observed, possibly suggesting a
type of spin glass-like behavior. The specific behavior uncovered was a
chaotic sequence of alternating strong and weak spin-spin correlations as
distance increased.

We suspect that the this behavior, if it carries over to spin glasses on
Euclidean lattices, arises from different physical origins than those
giving rise to chaotic size dependence.  Roughly speaking, the
renormalization-group studies of \cite{MB1,MB2} uncover how correlation
strengths change as one looks at spins increasingly farther apart.  In
contrast, chaotic size dependence focuses on a {\it fixed\/} correlation
function, for example $\langle\sigma_0\sigma_1\rangle$, calculated for each
$\Lambda_L$ using an ordinary Gibbs measure with, say, periodic boundary
conditions.  If $\langle\sigma_0\sigma_1\rangle$ changes chaotically as $L$
grows, a clear signal for many pure states is provided.  If it does not,
then the system likely has no more than a single pair of pure states.

We're not aware of any studies attempting to correlate the presence of many
states with changes in $\langle\sigma_x\sigma_y\rangle$ as $|x-y|$
increases, as in \cite{MB1,MB2}, but we suspect that such chaoticity may be
present {\it regardless\/} of the multiplicity of pure states.

Another type of chaoticity in spin glasses is {\it chaotic temperature
dependence\/} (CTD). This has been the subject of numerous studies, and
again can be detected using Migdal-Kadanoff renormalization group
techniques \cite{BB87}.  Roughly speaking, temperature chaos refers to the
erratic behavior of correlations, upon changing temperature, on
lengthscales that diverge as the temperature increment goes to zero.  It
was predicted \cite{BM87,FH88,BB87} for the EA spin glass as a consequence
of the scaling/droplet {\it ansatz\/}, but seemed to be implied as well by
the RSB theory \cite{K89,NH93,R94}.  More recent numerical and analytical
work (see \cite{BM99} and references therein) have led to claims that
chaotic temperature dependence is not present in either the SK or the EA
model (see also \cite{BDHV02}), although
\cite{BM02} allows the possibility of a weak effect at large lattice sizes.
Other work \cite{RC02} claims to see a very small effect at ninth order in
perturbation theory.  At this time the issue remains unresolved.

Chaotic temperature dependence and chaotic size dependence are clearly
different, given that the latter is seen at {\it fixed\/} temperature.
Nevertheless, the intriguing question arises: are the two somehow related?
We do not know the answer at this time.  It seems at least plausible that a
chaotic change in a fixed correlation function as the volume increases
(CSD) would imply a similar chaoticity as temperature changes (CTD), given
that the former corresponds to surface changes and the latter to bulk
changes.  On the other hand, chaotic size dependence only occurs (for, say,
periodic boundary conditions) when there exist many competing pure states,
and is absent for a single pair, as in droplet/scaling.  At the same time,
the droplet/scaling picture seems to require chaotic temperature
dependence, which arises there from a lack of `rigidity' (compared to a
ferromagnet) in the spin glass phase.  So at the very least, chaotic
temperature dependence does not seem to imply chaotic size dependence.
Whether the converse is true remains an open question.

\subsection{Effects of a Magnetic Field}
\label{subsec:magnetic}

In Sec.~\ref{subsec:SK} the stability analysis of deAlmeida and Thouless
\cite{AT78} was discussed.  Although the original intent was to study the
stability of the replica-symmetric SK solution \cite{SK75} in the $T$-$h$
plane, the consequences of the analysis of \cite{AT78} remain important for
short-ranged models as well.  In particular, the phase separation boundary
between the paramagnetic and spin glass phases in the $T$-$h$ plane is now
generically referred to as an ``AT'' line, and debate over its existence
for realistic spin glasses remains strong.  In the RSB picture, the AT line
begins at $(T_c,0)$ and extends upward through nonzero values of $h$ (see,
for example, Fig.~48 of \cite{BY86}).  In contrast, the droplet/scaling
model predicts that the spin glass phase is unstable to {\it any\/}
external magnetic field, no matter so small, resulting in no AT line.

However, in the droplet/scaling picture even a small magnetic field will
have dramatic dynamical consequences.  The droplet theory predicts a
``magnetic correlation length'' $\xi_h$ that diverges as an inverse power
of the field as $h\to 0$ (the power itself being a function of the domain
wall energy exponent $\theta$ defined in Sec.~\ref{subsubsec:free}).  The
magnetic correlation length describes roughly the lengthscale over which a
field $h$ will destroy local correlations in a spin glass phase.  On this
lengthscale, the characteristic relaxation time $\tau_{\xi_h}$ grows
exponentially with $\xi_h$ \cite{FH88}.  Consequently, spin glass
correlations can persist for unmeasurably long times in small fields.  This
makes it difficult to establish for a real spin glass sample whether a true
thermodynamic AT line exists or whether one is instead observing a
nonequilibrium dynamical effect.  In the droplet/scaling picture, one
therefore replaces the equilibrium AT line with a dynamical one, separating
regions in the $T$-$h$ plane where the system falls out of equilibrium on
accessible laboratory time scales (cf.~Fig.~3 in \cite{FH88}).  This
feature, however, can create difficulties in the interpretation not only of
laboratory experiments but also of numerical simulations (see, for example,
\cite{SH91,GH91}).

We turn now to our own results.  We point out first that all of our
theorems regarding self-averaging of overlaps, and therefore lack of
viability of the RSB picture in short-ranged models, are not affected by
addition of a magnetic field.  However, our results do {\it not\/} rule out
the presence of a true, thermodynamic AT line.  In particular, the
many-state chaotic pairs picture remains perfectly consistent with the
presence of such a line.  Whether an AT line ultimately exists will almost
certainly depend on the resolution of the problem of the number of states
in {\it zero\/} external field: in all likelihood, there is no line if only
a single pair of pure states exists, but it should be present if many
states exist.

       \section{Other Topics}
       \label{sec:other}

       \subsection{Metastable States}
       \label{subsec:metastable}

       Our analysis has focused largely on pure state structure and ordering in
       short-ranged spin glass models.  While our emphasis has centered on
       equilibrium thermodynamics, we have argued elsewhere \cite{NS99b} that pure
       states also deeply influence nonequilibrium dynamical phenomena, such as
       coarsening, aging, and others related to dynamical evolution following a
       deep quench.

       Another prominent, and much studied, feature of spin glasses is the
       presence of many {\it metastable\/} states, i.e., states that are
       energetically stable up to $M$-spin flips for some finite $M$, but unstable
       to flips of clusters of more than $M$ spins.  Colloquially speaking, these
       are spin configurations that are `local' rather than `global' energy (or
       free energy, at positive temperature) minima; their confining barriers are
       of height $O(1)$ rather than $O(N^\delta)$ for some $\delta>0$.  These states are
       believed to be responsible for much of the anomalously slow relaxation
       features of spin glasses \cite{DMH79,BK83}, and their presence has led to
       new numerical techniques such as simulated annealing \cite{KGV83,C85}.

       In \cite{NS99a}, we studied the properties of metastable states not only in
       EA spin glasses with continuous coupling distributions such as Gaussian,
       but also disordered ferromagnets.  We will confine ourselves here to spin
       glasses, and will only present our own results; the reader is referred to
       \cite{NS99a} for a more detailed discussion, as well as references to
       other work on metastability.

       To clarify the discussion, we define a 1-spin-flip stable state as an
       infinite-volume spin configuration whose energy, as given by
       Eq.~(\ref{eq:EA}), cannot be lowered by the flip of any single spin.
       Similarly, an $M$-spin-flip stable state ($M<\infty$) is an infinite-volume
       spin configuration whose energy cannot be lowered by the flip of any
       cluster of $1,2,\ldots,M$ spins.  Recall (Sec.~\ref{subsec:pure}) that a
       {\it ground\/} state is an infinite-volume spin configuration whose energy
       cannot be lowered by the flip of {\it any\/} finite cluster of spins (i.e.,
       $M\to\infty$).

       \subsubsection{A New Dynamical Method}
       \label{subsubsec:method}

       Our approach to studying metastable states is rather unusual, and based on
       a new dynamical technique: we construct a natural ensemble (the
       ``$M$-stable ensemble'') on states that evolve from an initial spin
       configuration generated through a deep quench via a zero-temperature
       ``lattice animal'' dynamics.

       We briefly describe this method, starting with the single-spin-flip
       case. Let $\sigma^0$ denote the initial (time zero) infinite-volume spin
       configuration on $Z^d$.  It is chosen from the infinite-temperature
       ensemble in which each spin is equally likely to be $+1$ or $-1$,
       independently of the others.  The continuous-time dynamics is given by
       independent, rate-1 Poisson processes at each site $x$ corresponding to
       those times $t$ at which the spin at $x$ looks at its neighbors and
       determines whether to flip.  It does so only if a flip lowers the system
       energy (that is, we consider only zero-temperature dynamics).  We denote by
       $\omega_1$ a given realization of this zero-temperature single-spin flip
       dynamics; so a given realization $\omega_1$ would then consist of a
       collection of random times $t_{x,i}$ ($x\in Z^d$, $i=1,2,\ldots$) at every
       $x$ when spin flips for the spin $\sigma_x$ are considered.

       Given the Hamiltonian (\ref{eq:EA}) and a specific ${\cal J}$, $\sigma^0$,
       and $\omega_1$, a system will evolve towards a single well-defined spin
       configuration $\sigma^t$ at time $t$ (this uses the fact that the
       individual couplings come from a continuous distribution such as
       Gaussian).  It is important to note that these
       three realizations (coupling, initial spin configuration, and dynamics) are
       chosen independently of one another.  The continuous coupling distribution
       and zero-temperature dynamics together guarantee that the energy per spin
       $E(t)$ is always a monotonically decreasing function of time.

       We now consider {\it multiple\/}-spin flips, in which we allow rigid flips
       of all lattice animals (i.e., finite connected subsets of $Z^d$, not
       necessarily containing the origin) up to size $M$.  The case $M=1$ is the
       single-spin flip case just described; $M=2$ corresponds to the case where
       both single-spin flips and rigid flips of all nearest-neighbor pairs of
       spins are allowed; and the case of general $M$ corresponds to flips of
       1-spin, 2-spin, 3-spin, $\ldots$ $M$-spin connected clusters.  A specific
       realization of this $M$-spin-flip dynamics will be denoted $\omega_M$.

       There is a technical issue that needs to be addressed, because we wish the
       dynamics to remain sensible even in the limit $M\to\infty$.  We require
       that the probability that any fixed spin considers a flip in a unit time
       interval remains of order one, uniformly in $M$.  Such a choice would
       guarantee, for example, that the probability that a spin considers a flip
       in a time interval $\Delta t$ vanishes as $\Delta t\to 0$, again uniformly
       in $M$.  A further requirement for the dynamics to be well-defined is that
       information not propagate arbitrarily fast throughout the lattice as $M$
       becomes arbitrarily large.  It is not difficult to construct such a
       dynamics, but we omit the technical details here; they can be found in
       \cite{NS99a}.

       \subsubsection{Results}
       \label{subsubsec:results}

       In this section we present some of the results found in \cite{NS99a}; we
       omit all proofs and detailed discussions.  Some of these results are
       expected, while others are surprising and at variance with the ``folk
       wisdom'' that has evolved over the years.  Aside from the intrinsic
       interest in the structure of metastability in spin glasses, we believe that
       the information obtained also sheds light on some aspects of pure state
       structure, if only by way of contrast.  All results given below are
       rigorous, and hold for a.e.~${\cal J}$, $\sigma^0$, and $\omega_M$.

       \medskip

       \noindent {\bf Existence and Number of Metastable States.} --- In an
       infinite system, the Hamiltonian~(\ref{eq:EA}) displays {\it uncountably\/}
       many $M$-spin-flip stable states, for all finite $M\ge 1$ and for all
       finite $d\ge 1$.

       \medskip

       \noindent {\bf Convergence of Dynamics Following a Deep Quench}. --- It is
       not obvious {\it a priori\/} whether the system evolves towards a single,
       final metastable state $\sigma^\infty$.  It can be proved, however, that
       such a final state exists almost surely.  Equivalently, every spin flips
       only finitely many times.  (This is in contrast to, say, the $2D$ Ising
       ferromagnet, where every spin flips {\it infinitely\/} many times
       \cite{NS99b}.)

       \medskip

       \noindent {\bf How Much Information is Contained in the Initial State?}
       --- For $M=1$ in $1D$, precisely half the spins in $\sigma^\infty$ are
       completely determined by $\sigma^0$, with the other half completely
       undetermined by $\sigma^0$.  For higher $d$ and the same dynamics, it can
       be shown that a dynamical order parameter $q_D$, measuring the percentage
       dependence of $\sigma^\infty$ on $\sigma^0$, is strictly between 0 and 1.

       \medskip

       \noindent {\bf Size of Basins of Attraction} --- The basins of attraction
       of the individual metastable states are of negligible size, in the
       following sense: almost every $\sigma^0$ is on a boundary between two or
       more metastable states.  Equivalently, the union of the domains of
       attraction of {\it all\/} of the metastable states forms a set of measure
       zero, in the space of all $\sigma^0$'s.  (A similar result for pure states
       was proved in \cite{NS99b}.)

       \medskip

       \noindent {\bf Distribution of Energy Densities.} --- For any $M$, almost
       every $M$-spin-flip stable state has the same energy density, $E_M$.
       Moreover, the dynamics can be chosen so that $E_1>E_2>E_3>\ldots$ , and
       furthermore $E_M$ for any finite $M$ is larger than the ground state energy
       density, which is the limit of $E_M$ as $M \to \infty$.

       \medskip

       \noindent {\bf Overlaps.} --- Almost every pair of metastable states,
       whether two $M$-spin-flip stable states or one $M$- and one $M'$-spin-flip
       stable state, has zero spin overlap.  So the overlap distribution function,
       for either {\it all\/} of the metastable states, or only all those with any
       fixed $M$, is simply a $\delta$-function at zero.

       \medskip

       \noindent {\bf Scaling of Number of Metastable States with Volume.} ---
       For sufficiently large volumes, the number of metastable states in finite
       samples scales exponentially with the volume in general $d$ for states of
       any stability.

       \medskip

       \noindent {\bf Remanent Magnetization.} --- If the initial state is
       uniformly $+1$ in $1D$, the remanent magnetization is known to be $1/3$
       (for $M=1$)\cite{FM79}. In higher dimensions, a heuristic calculation for the Gaussian
       spin glass gives a rigorous lower bound on the remanent magnetization that for large
       $d$ behaves like $e^{-2d\log(d)}$.  Exact results can be obtained in all
       $d$ for some other models \cite{NS99a}.

       \medskip

       \noindent {\bf Correspondence Between Pure and Metastable States.} --- More
       precisely, if there are multiple pure or ground state pairs, does the spin
       configuration corresponding to a typical metastable state ``live in'' the
       domain of attraction of a single pure or ground state, as is commonly
       thought?  The answer is {\it no\/} in both cases: almost every metastable
       state will be on a boundary in configuration space between multiple pure
       or ground states.

       \bigskip

       Finally, we ask: does knowledge of metastable states provide information on
       the number or structure of thermodynamic ground or pure states?  So far,
       the answer seems to be largely no.  For example, in the one-dimensional
       spin glass there is only a single pair of thermodynamic ground states, but
       an uncountable number of infinite-volume $M$-spin-flip stable states for
       any finite $M$.  This example illustrates a potential difficulty with
       numerical studies in higher dimensions, aimed at determining the number of
       ground or pure states: the presence of many metastable states could
       complicate interpretations of these studies.

       \subsection{The Statistical Mechanics of Homogeneous vs.~Disordered Systems}
       \label{subsec:statmech}

       Our hope is that this topical review has conveyed some of the depth and
       richness of the physics and mathematics of the equilibrium statistical
       mechanics of spin glasses.  We conclude by returning briefly to Question 7,
       raised in Sec.~\ref{sec:open}: In what ways do we now understand how the
       statistical mechanical treatment of systems with quenched disorder differs
       in fundamental ways from that of homogeneous systems?  In some sense, much
       of this review focused on one or another aspect of this question, but it
       may be helpful for us to tie together some of the common threads that have
       run through much of this review.  We emphasize that we will not attempt to
       provide a comprehensive or even an extensive answer to this question, which
       remains largely open, but will only focus on some of the insights that our
       analysis of the spin glass problem may have uncovered.  (A similar
       discussion appears also in \cite{St03}.)

       Most homogeneous systems that can be treated by classical equilibrium
       statistical mechanics share several salient features.  Whether one is
       studying crystals, ferromagnets, ferroelectrics, liquid crystals, or even
       some quantum systems such as superconductors and superfluids, the analysis
       of the low-temperature phase is simplified by various {\it spatial\/}
       symmetries, such as translational, orientational, gauge, and others.  Of
       course, no such symmetries are manifestly apparent in many disordered
       systems.

       It has also been known for many years \cite{Br59} that averaging over the
       quenched disorder presents additional complications.  But this issue has
       been extensively discussed in many other reviews
       \cite{BY86,Chowd86,FH91,BCKM98,MPV87,Stein89,Dotsenko01} and will not be
       further addressed here.  Another complicating feature present in some (but
       not all) disordered systems is frustration.  While frustration may also
       occur in homogeneous systems (e.g., triangular antiferromagnets), its joint
       presence with quenched disorder may result in more profound effects.  We
       hereafter confine ourselves to systems with both disorder {\it and\/}
       frustration, taking the spin glass as their prototype.

       The nature of the low-temperature phase is a clear point of departure
       between homogeneous systems and spin glasses (assuming they have a
       low-temperature phase).  Homogeneous systems typically display a relatively
       simple order parameter, representing the nature of ordering in a pure or
       ground state that is unique up to an overall simple symmetry transformation
       that leaves the Hamiltonian invariant.  It remains unknown whether the spin
       glass similarly possesses a simple order parameter in the EA model in
       finite dimensions, but it certainly does not in the infinite-ranged SK
       model.

       The striking contrast between the nature of broken symmetry in the RSB
       theory, as opposed to that present in most homogeneous systems, is clearly
       an important difference between them and infinite-ranged spin glasses.
       However, as extensively discussed in this review, it now appears that this
       type of broken symmetry is absent in short-ranged spin glass models.

       However, rather than seeing this as a disappointing piece of news, our
       viewpoint is that it is perhaps indicative of a far deeper contrast between
       spin glasses and simple homogeneous systems: in all such latter systems of
       which we are aware, mean field theory has been invaluable in providing the
       basic information concerning the nature of broken symmetry, order
       parameter(s), and low-temperature behavior in general.  Mean field theory's
       main shortcoming lies in its behavior very close to the transition
       temperature, but even this failure usually occurs only in sufficiently low
       dimensions (which often includes the physically interesting case of $3D$):
       there usually exists an upper critical dimension above which mean field
       theory also provides the correct critical exponents.

       If the infinite-ranged SK model becomes exact for the EA model in
       infinite dimensions, as is commonly believed, then we have a new
       feature whose contrast with homogeneous systems is perhaps even more
       striking than the presence of RSB: that is, the failure of mean
       field theory to provide a correct description of the low-temperature
       phase, even far from the critical point, in {\it any\/} finite
       dimension.  Equivalently, the $d\to\infty$ limit of the EA model is
       singular.  This possibility was broached by Fisher and Huse
       \cite{FH87b}, and our work confirms their conjecture.  Some of the
       reasons why mean field theory fails in any finite dimension are
       presented in Sec.~\ref{subsec:csdsk} and reviewed in
       Sec.~\ref{subsec:summary}.

It is important to emphasize that the disconnect between the SK and EA
models is profound \cite{NS03}: when one tries to transfer concepts in {\it
either\/} direction, contradictory and even absurd results can ensue.  One
interesting question discussed in \cite{NS03}, worth repeating here, is
whether a new type of thermodynamic object in place of states may be
appropriate for understanding the meaning of replica symmetry breaking in
the SK model.  The usual notions of states are local ones: that is, they
describe the behavior of correlation functions (at fixed locations) and
related quantities.  These notions do not seem to work for the SK model.
This suggests the intriguing possibility that a more global object might be
constructible that would provide a more natural `fit' for the SK model.  We
do not know whether this will turn out to be the case, but it is clear that
if it does, such an object would be substantially different from the
thermodynamic states that have been used up until now. 

       A different issue concerns the nonexistence of a thermodynamic limit
       for states (or equivalently, correlations).  This is manifested as
       chaotic size dependence, and occurs when coupling-independent
       boundary conditions are used and there exist many observable pure
       states.  This is really a reflection of the lack of any spatial
       symmetries that allow one to choose boundary conditions, or an
       external symmetry-breaking field, that can lead to the existence of
       such a limit.

       One could, of course, artificially obtain chaotic size dependence even in a
       simple homogeneous ferromagnet below $T_c$ by, say, choosing random
       boundary conditions independently in each $\Lambda_L$.  But there one also
       knows how to choose boundary conditions to obtain a limiting pure state,
       including the interface states, whose analogues would be invisible pure
       states in spin glasses (Sec.~\ref{sec:interfaces}).  The option of choosing
       boundary conditions that lead to a limiting pure state does not now exist
       for spin glasses if they possess many pure state pairs; and there may or
       may not be fundamental reasons preventing that option from ever becoming
       available.  Regardless of whether it does, it remains interesting and
       useful that the presence of chaotic size dependence provides a clear signal
       of the existence of many states (Sec.~\ref{sec:csd}).

Besides chaotic size dependence, it has been speculated that spin glasses,
both short-ranged and infinite-ranged, display {\it chaotic temperature
dependence\/}, as discussed in Sec.~\ref{subsubsec:chaos}.  As indicated
there, whether chaotic temperature dependence actually exists in spin
glasses remains an open question.  However, its potential presence in spin
glasses represents a qualitatively new thermodynamic feature of at least
some types of disordered systems.

       A further contrast between homogeneous and disordered systems is provided
       by the presence of observability vs.~invisibility of different types of
       pure states.  This is intimately interwoven with chaotic size dependence,
       and also with the nature of the interfaces separating these states; these
       interfaces can, in principle, be different from those seen in homogeneous
       systems (Sec.~\ref{sec:interfaces}).  These issues all arise from the use
       of coupling-dependent vs.~independent boundary conditions.  No analogue for
       this distinction exists in homogeneous systems, but it is very possibly of
       basic importance in systems with quenched disorder.

       Finally, we would like to emphasize that if this entire discussion has
       uncovered any fundamental unifying principle, it is the set of ideas and
       techniques encapsulated in the construct of the {\it metastate.\/} It is
       our strong belief that any final understanding of the spin glass phase, and
       possibly that of other inhomogeneous systems, will make extensive use of
       this concept.
	
\bigskip

{\it Acknowledgments.\/} We thank T.~Rosenbaum for suggesting we write this
review, and A.~van~Enter and P.~Contucci for useful remarks on the
manuscript.  We are also indebted to M.A.~Moore for a number of
illuminating comments and for pointing us towards several useful
references.

	\renewcommand{\baselinestretch}{1.0}


\begin{thebibliography}{10}
\bibitem{noteground} The term `ground state' is also sometimes used more
broadly, to denote measures on multiple ground state configurations.  In
this review, we avoid this terminology, and use the term `ground state' to
refer only to a single, infinite-volume, spin configuration with the
properties described above.
\end{thebibliography}
	\end{document}